\documentclass[twoside,twocolumn]{article}
\usepackage[super,sort&compress,comma]{natbib}
\usepackage[version=3]{mhchem}
\usepackage{times,mathptmx}
\usepackage{sectsty}
\usepackage{balance}
\usepackage{amssymb}
\usepackage{color}
\usepackage{graphicx} 
\usepackage{lastpage}
\usepackage{fancyhdr}
\usepackage{bm}
\usepackage{float}
\usepackage{array}
\usepackage{textcomp}

\graphicspath{{./Figures/}{./}}

\begin{document}

\thispagestyle{plain}
\fancypagestyle{plain}{
\renewcommand{\headrulewidth}{1pt}}
\renewcommand{\thefootnote}{\fnsymbol{footnote}}
\renewcommand\footnoterule{\vspace*{1pt}%
\hrule width 3.4in height 0.4pt \vspace*{5pt}}
\setcounter{secnumdepth}{5}

\makeatletter
\def\subsubsection{\@startsection{subsubsection}{3}{10pt}{-1.25ex plus -1ex minus -.1ex}{0ex plus 0ex}{\normalsize\bf}}
\def\paragraph{\@startsection{paragraph}{4}{10pt}{-1.25ex plus -1ex minus -.1ex}{0ex plus 0ex}{\normalsize\textit}}
\renewcommand\@biblabel[1]{#1}
\renewcommand\@makefntext[1]%
{\noindent\makebox[0pt][r]{\@thefnmark\,}#1}
\makeatother
\renewcommand{\figurename}{\small{Fig.}~}
\sectionfont{\large}
\subsectionfont{\normalsize}

\fancyfoot{}
\fancyhead{}
\renewcommand{\headrulewidth}{1pt}
\renewcommand{\footrulewidth}{1pt}
\setlength{\arrayrulewidth}{1pt}
\setlength{\columnsep}{6.5mm}
\setlength\bibsep{1pt}

\twocolumn[
  \begin{@twocolumnfalse}
\noindent\LARGE{\textbf{Effect of fluid-colloid interactions on the mobility of a thermophoretic microswimmer in non-ideal fluids}} 
\vspace{0.6cm}

\noindent\large{\textbf{Dmitry A. Fedosov$^a$, Ankush Sengupta$^a$, and Gerhard Gompper$^a$}}
\vspace{0.5cm}

\noindent 
\normalsize{
Janus colloids propelled by light, e.g., thermophoretic particles, offer promising prospects as artificial 
microswimmers. However, their swimming behavior and its dependence on fluid properties and fluid-colloid 
interactions remain poorly understood. Here, we investigate the behavior of a thermophoretic Janus 
colloid in its own temperature gradient using numerical simulations. The dissipative particle dynamics
method with energy conservation is used to investigate the behavior in non-ideal and ideal-gas 
like fluids for different fluid-colloid interactions, boundary conditions, and temperature-controlling 
strategies. The fluid-colloid interactions appear to have a strong effect on the colloid behavior, since
they directly affect heat exchange between the colloid surface and the fluid. The simulation results show 
that a reduction of the heat exchange at the fluid-colloid interface leads to an enhancement of 
colloid's thermophoretic mobility. The colloid behavior is found to be different in non-ideal and ideal 
fluids, suggesting that fluid compressibility plays a significant role. The flow field around the colloid 
surface is found to be dominated by a source-dipole, in agreement with the recent theoretical and 
simulation predictions. Finally, different temperature-control strategies do not appear to have a strong 
effect on the colloid's swimming velocity.

 \vspace{0.5cm}}
 \end{@twocolumnfalse}

]



\footnotetext{\textit{$^{a}$~Theoretical Soft Matter and Biophysics,
Institute of Complex Systems and Institute for Advanced Simulation,
Forschungszentrum J\"ulich, 52425 J\"ulich, Germany;  Email: d.fedosov@fz-juelich.de
}}


\section{Introduction}

The construction of nano- and micro-machines, which can move through a 
fluid environment, is one of the grand challenges confronting nanoscience 
today \cite{Jones_SM_2004,Ozin_DNM_2005,Paxton_CLM_2006,Ebbens_PNS_2010}. 
Several strategies and physical mechanisms have been
employed so far to generate self-propulsion in a fluid \cite{Lauga_HSO_2009,Elgeti_PMS_2015}. One approach is 
biomimetic, where the flagellar propulsion of sperm,
bacteria, or cilia is recreated with synthetic soft materials and 
actuators. Some recent examples include artificial sperm \cite{Williams_SPS_2014} and
artificial cilia \cite{Sareh_SLA_2013}; however, in most of these cases, the 
machines are rather of millimeter than of sub-micrometer size. Nano- to micrometer
length scales are reached by artificial cilia made from microtubules and 
motorproteins \cite{Sanchez_CLB_2011},
and by magnetic nano- and microscrews rotated by an external magnetic field 
\cite{Schamel_NPA_2014}.
Another approach is physico-chemical, where non-equilibrium concentration fields
or temperature distributions in the fluid environment are generated around the 
swimmer and are employed for propulsion, without any movable parts of the swimmer itself.  
Here, diffusiophoretic 
\cite{Golestanian_PMM_2005,Howse_SMP_2007,Rueckner_CPD_2007,Thakur_DSP_2011,Popescu_PMS_2010,Palacci_LCL_2013} 
and thermophoretic 
\cite{Jiang_AMP_2010,Rings_HBM_2010,Golestanian_DFV_2010,Yang_TNS_2011,Bickel_FPV_2013,deBuyl_PSP_2013} 
Janus colloids have been studied most intensively. For diffusiophoretic swimmers, a 
semi-spherical cap on the colloidal Janus particle catalyzes a
reaction in the fluid and thereby generates a spatially inhomogeneous non-equilibrium 
distribution of reaction agents and products. For thermophoresis, a semi-spherical cap 
on a colloidal Janus particle absorbs light from an external light source, and is thereby
heated, and generates a local temperature gradient. An interesting combination of the
two has also been investigated, where the onset of the chemical reaction is triggered
by the external light intensity \cite{Palacci_LCL_2013}. Light-controlled microswimmers have
the advantage that their motion can be controlled easily by a variation of the light
intensity.

A particularly interesting type of thermophoretic microswimmers has been suggested by 
Volpe et al. \cite{Volpe_MSE_2011}. This is again a colloidal particle with a metallic,
light-absorbing cap; however, this Janus colloid is immersed in a binary fluid mixture
at an ambient temperature just below its lower demixing critical point. A slight heating
of the cap then leads to a local demixing of the fluid mixture, which generates the
driving force for swimming. An advantage of this mechanism is that it works for much
smaller power of the light source than that for thermophoretic microswimmers in single-component
fluids.

The study of thermophoretic swimming in binary fluids \cite{Volpe_MSE_2011} demonstrates 
that the interaction of the fluid with the colloid surface plays an important role. 
In particular, it was shown by Volpe et al. \cite{Volpe_MSE_2011} that the swimming
direction depends on which of the two components partially wets the colloid surface.
However, theoretical studies have only considered either fluids on the
level of the incompressible Navier-Stokes equation \cite{Bickel_FPV_2013}, or fluids 
with an ideal-gas equation of state \cite{Yang_TNS_2011,deBuyl_PSP_2013,Yang_HSS_2014}. 
Thus, as a first step toward an understanding of
thermophoretic self-propulsion in real fluid mixtures, we investigate a system with
a single-component fluid with a non-ideal equation of state and a variety of boundary
conditions (BCs) on the colloid surface. In particular, we investigate the influence 
of no-slip and slip BCs and fluid-colloid interactions. In addition, the behavior of 
a thermophoretic swimmer is studied for different temperature controls and gradients. 
We find that local fluid-colloid interactions and temperature gradients near the colloid's 
surface control its swimming velocity. The models with non-ideal and ideal fluids lead 
to qualitatively different trends in the colloid mobility. Finally, the flow field around 
a swimming thermophoretic colloid is mainly determined by the source-dipole term in agreement 
with recent theoretical predictions \cite{Bickel_FPV_2013}.

\section{Models and methods}

We consider a spherical Janus colloid immersed in a single-component fluid.
In order to correctly describe the hydrodynamics of a thermophoretic microswimmer,
the fluid model has to properly conserve momentum and energy locally.
We employ here the version of dissipative-particle-dynamics (DPD) approach 
\cite{Hoogerbrugge_SMH_1992,Espanol_SMO_1995} with energy conservation \cite{Espanol_DPDE_1997,Avalos_DPDE_1997}. 

\subsection{Dissipative particle dynamics with energy conservation (eDPD)}

In the standard (isothermal) DPD approach \cite{Hoogerbrugge_SMH_1992,Espanol_SMO_1995}, 
the fluid is modeled by a set of $N$ particles, each of mass $m$,
interacting through a weak conservative (${\bf F}^C_{ij}$),
a dissipative (${\bf F}^D_{ij}$), and random (${\bf F}^R_{ij}$) pairwise forces, 
where the subscripts $i, j$ refer to particle indices. 
The pairwise additive forces are given by
\begin{eqnarray}
\label{eq:dpd}
&& {\bf F}^C_{ij}= a \omega^C(r_{ij}){\bf \hat{r}}_{ij}, \nonumber\\
&& {\bf F}^D_{ij}=-\gamma \omega^D(r_{ij})({\bf v}_{ij} \cdot {\bf \hat{r}}_{ij}){\bf \hat{r}}_{ij}, \\
&& {\bf F}^R_{ij}=\sigma \omega^R(r_{ij}) \xi_{ij} \Delta t^{-1/2} {\bf \hat{r}}_{ij}, \nonumber
\end{eqnarray}
where ${\bf r}_{ij}={\bf r}_i-{\bf r}_j$, ${\bf \hat{r}}_{ij}={\bf r}_{ij}/r_{ij}$ is its unit vector, and 
${\bf v}_{ij}={\bf v}_i-{\bf v}_j$ with ${\bf r}$ and ${\bf v}$ being the particle positions and velocities,
respectively. The parameter $a$ is the conservative force coefficient, which affects fluid compressibility. 
$\gamma$ and $\sigma$ are the friction and noise amplitudes, which are related through the fluctuation-dissipation 
balance \cite{Espanol_SMO_1995} as $\sigma^2=2\gamma k_BT$, where 
$k_B$ is the Boltzmann constant and $T$ is the temperature. All forces are short-ranged with a cutoff 
radius $r_c$, and vanish for $r_{ij} > r_c$. The conservative force profile is defined by 
$\omega^C(r_{ij})= (1-r_{ij}/r_c)$ for $r_{ij}\leq r_c$, while the spatial dependence of the dissipative ($\omega^D$) 
and random ($\omega^R$) weight functions is determined by $\omega^R(r_{ij})= (1-r_{ij}/r_c)^s$ for $r_{ij}\leq r_c$ and 
the relation $\omega^D=(\omega^R)^2$ derived from the fluctuation-dissipation
theorem \cite{Espanol_SMO_1995}. Here, $s$ is the exponent which affects inter-particle friction and
fluid viscosity such that $s < 1$ is advantageous in order to achieve a sufficiently large fluid viscosity
\cite{Fan_DNA_2006,Fedosov_DLP_2008}. $\xi_{ij}$ is a Gaussian distributed random variable with
zero mean and unit variance with the requirement $\xi_{ij}=\xi_{ji}$ for momentum conservation. Finally, 
$\Delta t$ is the simulation time step.

The particle dynamics is determined by the equations of motion as 
\begin{equation}
\dot{{\bf r}}_i = {\bf v}_i, ~~~~~~~~~
\dot{{\bf v}}_i = \frac{1}{m}\sum_j ({\bf F}^C_{ij} + {\bf F}^D_{ij} + {\bf F}^R_{ij}),
\end{equation}
which are integrated using the velocity-Verlet algorithm \cite{Allen_CSL_1991}.

In the energy-conserving DPD (eDPD) method \cite{Espanol_DPDE_1997,Avalos_DPDE_1997}, each 
fluid particle $i$, in addition to its position and velocity, also possesses an internal energy 
$\epsilon_i$. We assume that the internal energy of a particle is related to the temperature 
value $T_i$ as $\epsilon_i = C_v T_i$, where $C_v$ is the specific heat of a fluid \cite{Espanol_DPDE_1997}. 
Then, the evolution equation for particle temperature can be written in the following
form  
\begin{equation}
C_v \dot{T_i} = q_i = \sum_j (q_{ij}^c + q_{ij}^w),
\end{equation}
where $q_i$ is the heat flux between particle $i$ and the neighboring particles $j$
within the cutoff radius $r_c$. The heat flux $q_i$ is a sum of pairwise contributions from 
the heat conduction ($q_{ij}^c$) due to local temperature gradients between particles and 
the work ($q_{ij}^w$) done by the conservative, dissipative, and random forces. For instance, 
the work done by the dissipative force corresponds to viscous heating.

Local heat conduction between particles is defined as \cite{Espanol_DPDE_1997,Avalos_DPDE_1997} 
\begin{equation}
q_{ij}^c = k_{ij} \omega_c^2(r_{ij})\left(\frac{1}{T_i} - \frac{1}{T_j} \right) + \alpha_{ij}
\omega_c(r_{ij})\zeta_{ij}\Delta t^{-1/2},
\label{eq:cond}
\end{equation}  
where $k_{ij}$ is the thermal conductivity coefficient, $\alpha_{ij}$ is the noise amplitude, 
and $\zeta_{ij}$ is the associated noise modeled from the Gaussian probability 
distribution with zero mean and unit variance and with the condition $\zeta_{ji}=-\zeta_{ij}$. Note that 
$q_{ij}^c$ in Eq. (\ref{eq:cond}) consists of deterministic and random heat-conduction terms 
and provides local energy conservation. The conductivity coefficients are defined as \cite{Espanol_DPDE_1997,Avalos_DPDE_1997} 
\begin{equation}
k_{ij}=\frac{k_0C_v^2}{4k_B}(T_i+T_j)^2,~~~~~~~  \alpha_{ij}^2 = 2k_B k_{ij},  
\end{equation} 
where $k_0$ is a nominal strength of interparticle heat conductivity. For simplicity, we also 
select $\omega_c(r_{ij})=\omega^R(r_{ij})$. 

To connect particle dynamics to its internal temperature (or energy) \cite{Espanol_DPDE_1997,Avalos_DPDE_1997}, 
the force coefficients in Eq. (\ref{eq:dpd}) become functions of temperature and should be replaced by the corresponding 
$a_{ij}(T_i,T_j)$, $\gamma_{ij}(T_i,T_j)$, and $\sigma_{ij}(T_i,T_j)$ coefficients. 
Recently, it has been suggested that the conservative force coefficient $a_{ij}$ should depend linearly on temperature in order 
to properly reproduce fluid compressibility \cite{Li_EDPD_2014}. However, in the current work we employ 
a constant conservative force coefficient such that $a_{ij}(T_i,T_j)=a_F$. For simplicity, we also assume
no temperature dependence of the random force coefficient such that $\sigma_{ij}(T_i,T_j)=\sigma$. Then, the corresponding 
friction coefficient is given by 
\begin{equation}
\gamma_{ij} = \frac{\sigma^2}{4k_B}\left(\frac{1}{T_i} + \frac{1}{T_j} \right).  
\end{equation}  
The above expression for the inter-particle friction also determines the fluid's viscosity, which is controlled by setting 
the parameter $\sigma$ in simulations.  

The work fluxes $q_{ij}^w$ can be derived from the total energy $E$ of a simulated system. The total energy 
of the system, $E = \sum_i \epsilon_i + E_{mech}$, should be constant and thus, a change in mechanical energy $E_{mech}$ 
should correspond to the change in the internal energy such that $dE_{mech}=-d\left(\sum_i \epsilon_i\right)$. The mechanical 
energy consists of kinetic and potential contributions given by 
\begin{equation}
E_{mech} = \sum_i \frac{m v_i^2}{2} + \sum_{i \ne j} \frac{\phi(r_{ij})}{2},
\label{eq:emech}
\end{equation} 
where $\phi(r_{ij})$ is the interaction potential giving rise to the conservative force ${\bf F}^C_{ij} (r_{ij}) 
= - \nabla \phi(r_{ij})$. Therefore, we obtain $\phi(r) = 0.5a r_c(1-r/r_c)^2$. One way \cite{Espanol_DPDE_1997,Avalos_DPDE_1997} 
to derive the work fluxes $q_{ij}^w$ is to take the differential $dE_{mech}$ from the mechanical energy 
in Eq. (\ref{eq:emech}) and substitute the corresponding terms with the force expressions from Eq. (\ref{eq:dpd}) 
using the equations of motion. Such a method does not strictly provide energy conservation, and therefore, 
relatively small time steps might be required to have acceptably small energy variations. Another method is to 
implement energy conservation locally without explicit calculation of the $q_{ij}^w$ fluxes. Here, we assume 
that a change in the internal energy locally is equal to the change in both kinetic and potential energies such 
that $d\epsilon_i= - d(mv_i^2)/2 - d\left( \sum_j \phi(r_{ij})/2 \right)$. Thus, we calculate the changes in 
kinetic and potential energies locally after each integration step and adjust the internal energy of each particle 
accordingly. This method leads to exact conservation of the total energy of a system, and we have verified that it 
properly represents temperature gradients in a fluid. A similar idea has been used in Ref. \cite{Lisal_DPD_2011}, 
where energy conservation has been implemented locally for each pair of interacting fluid particles.

\subsection{Janus colloid and boundary conditions}
\label{sec:model}

\begin{figure}[h!]
\centering
\includegraphics[scale=0.35]{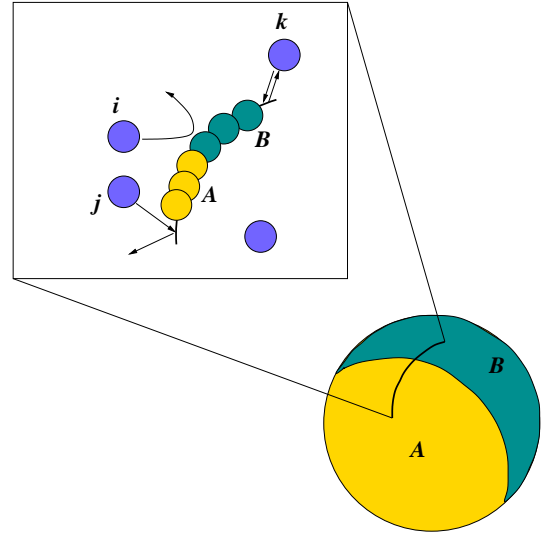}
\caption{Schematic diagram of the Janus colloid with the $A$ and $B$ caps. The inset
is a zoomed part of the cap along an arc showing two types of immobile particles
which constitute the colloid surface. The cap particles (type $A$ and type $B$)
shown in different colors interact with nearby fluid particles (e.g. $i$)
through the soft short-ranged DPD interactions, which are similar to those between 
fluid particles (shown in blue). In addition to the DPD interactions, bounce-back and 
specular reflection boundary conditions are used (here shown in the same picture, for convenience)
at the colloid-fluid boundary represented by the arc. For specular reflection, a fluid particle, 
e.g. $j$, will be reflected (shown by the arrows) at the spherical surface such that the tangential
velocity component is unaffected and the normal component is reversed. For the case of bounce-back 
condition, a fluid particle, e.g. $k$, will be reflected back along its incident direction 
as shown by the arrows.}
\label{fig:colloid}
\end{figure}

The Janus colloid is modeled with $N_s = 4000$ DPD particles placed on the
surface of a sphere with radius $R_s = 4r_c$ corresponding to the surface density of 
$\rho_{s} \approx 20/r_c^2$. The colloid is centered at the origin, with the hot (cold) side positioned 
in the half-space $x<0$ ($x>0$). The colloid particles are frozen at their positions 
on the spherical surface, see Fig.~\ref{fig:colloid}. It is important to note that a colloid moving 
in a resting fluid and a colloid fixed at a position with certain orientation are essentially identical; 
the main difference is just a transformation of the reference frame. In the former case, the colloid is moving 
in the fluid, while in the latter case the fluid is moving around the colloid. The two cases are equivalent 
for small rotational and translational diffusion coefficients, i.e. for sufficiently large colloids. There might 
be deviations for small colloids, when the relaxation time of the temperature profile becomes comparable to
the rotational diffusion time.
 
In the simulations, one or both sides of the colloid surface are maintained at constant
temperature. The steady-state behavior of the system is ensured by the constancy of the temperature gradient
in the fluid and of the flow field in the co-moving frame of the colloid. The fluid particles are present both 
outside and inside the colloidal shell, in order to ensure a proper pressure balance on both sides of
the colloid surface. Thus, the fluid particles inside and outside the colloid interact through the conservative 
force in Eq. (\ref{eq:dpd}). However, the friction coefficient between the inside and outside fluids is 
set to zero, since viscous interactions between them are shielded by the colloid wall. This means that 
$\gamma_{ij} = \sigma_{ij} = 0$ whenever $(i,j)$ represents a pair of fluid particles intercepted by the colloid 
surface.

Two separate cases of boundary conditions (BCs) are considered including either bounce-back collisions or specular reflection
of the fluid particles at the fluid-colloid interface. This applies to collisions both at the inside and the outside of 
the colloidal shell. The reflections of fluid particles are necessary to prevent the entry of a fluid particle from the 
exterior to the interior of the colloidal shell, or vice versa, since the conservative interactions between fluid and 
colloid particles are too soft to guarantee no inter-fluid mixing. On the other hand, the two different collision rules
realize different BCs at the colloid surface. The bounce-back collisions implement a no-slip (or stick) 
BCs, since the particle velocity is inverted at the colloid surface (i.e. ${\bf v} \to {\bf -v}$) resulting on average
in a vanishing tangential component of the fluid velocity. In the case of specular reflections, the velocity component 
parallel to the local tangent plane of the colloid surface remains unchanged, while the perpendicular component is inverted. 
This realizes slip BCs. 

In addition to the above mentioned BCs, the fluid particles also interact pairwise with the immobile DPD 
particles comprising the colloid surface. Therefore, different halves of the Janus particle possess not only 
dissimilar thermal properties, but also may have different fluid-colloid interactions (see Fig.~\ref{fig:colloid}). 
This is done by assigning different pair-interaction coefficients for the particles at the two halves of the colloid 
with the surrounding fluid particles. The particles of the two distinct hemispheres are labeled as $A$ and $B$. 
Given that the fluid particles $i$ and $j$ interact with a coefficient $a_{ij}=a_F$, the coefficients for fluid-colloid 
interactions will be referred to as $a_{iA}$ and $a_{iB}$ for the two halves, respectively. Then, $a_{iA} < a_F$  
mimics an effectively ``solvophilic'' surface, while $a_{iA} > a_F$ mimics an effectively ``solvophobic'' surface.
In experiments, Janus particles can be chemically functionalized to generate hydrophilic or hydrophobic interaction 
with the host fluid \cite{Chen_CJS_2011}. Such chemical functionalization can affect the particle behavior in addition to the 
temperature gradient. Thus, changing the fluid-cap interaction strength for both caps $A$ and $B$ allows us to access 
surface-tuning capabilities together with the thermophoretic control in our model. The dissipative and random force 
coefficients for fluid-colloid interactions are set to zero ($\gamma_{iA} = \sigma_{iA} = \gamma_{iB} = \sigma_{iB} = 0$), 
since the bounce-back and specular reflections of particles
at the colloid interface already define the type of BCs employed. Finally, the fluid and the colloid exchange heat 
locally within $r_c$ following Eq. (\ref{eq:cond}). In order that the fluid particles can approach the colloid particles 
close enough and exchange heat, a shorter cut-off radius for the fluid-colloid conservative interactions, 
${r^{\prime}}_c = 0.25 r_c$, has been used.   

\begin{figure}[h!]
\centering
\includegraphics[scale=0.4]{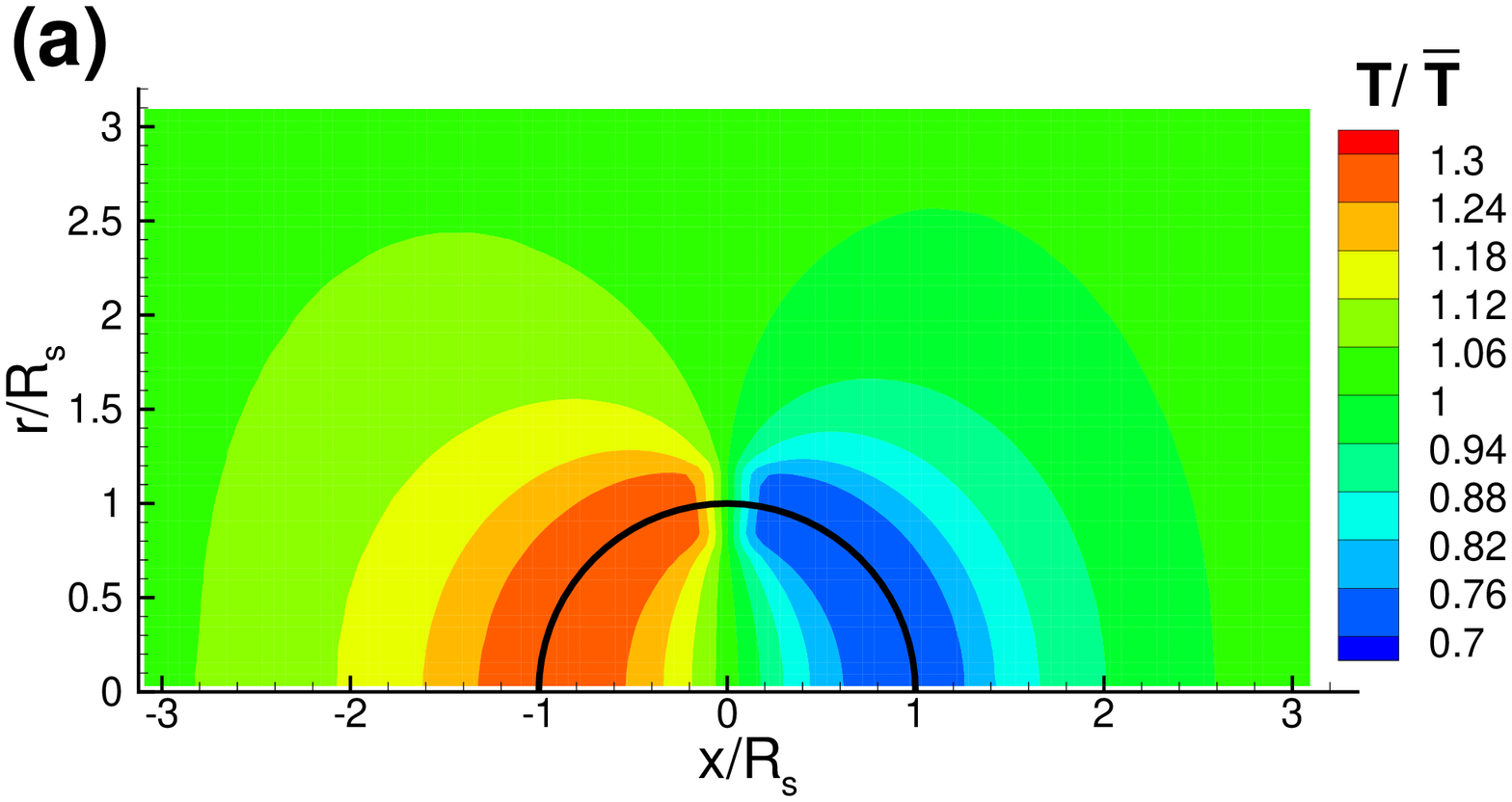}
\includegraphics[scale=0.4]{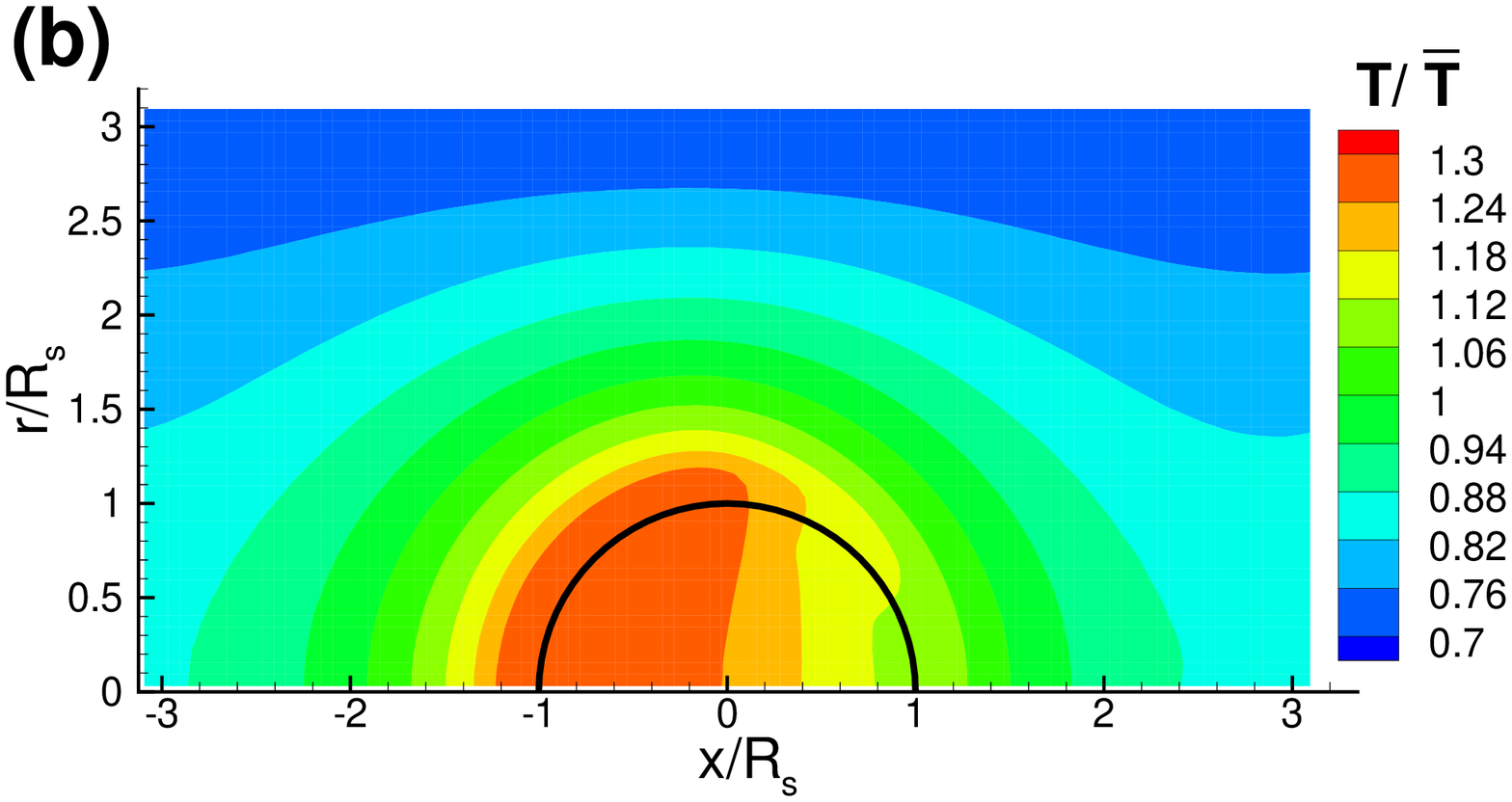}
\includegraphics[scale=0.4]{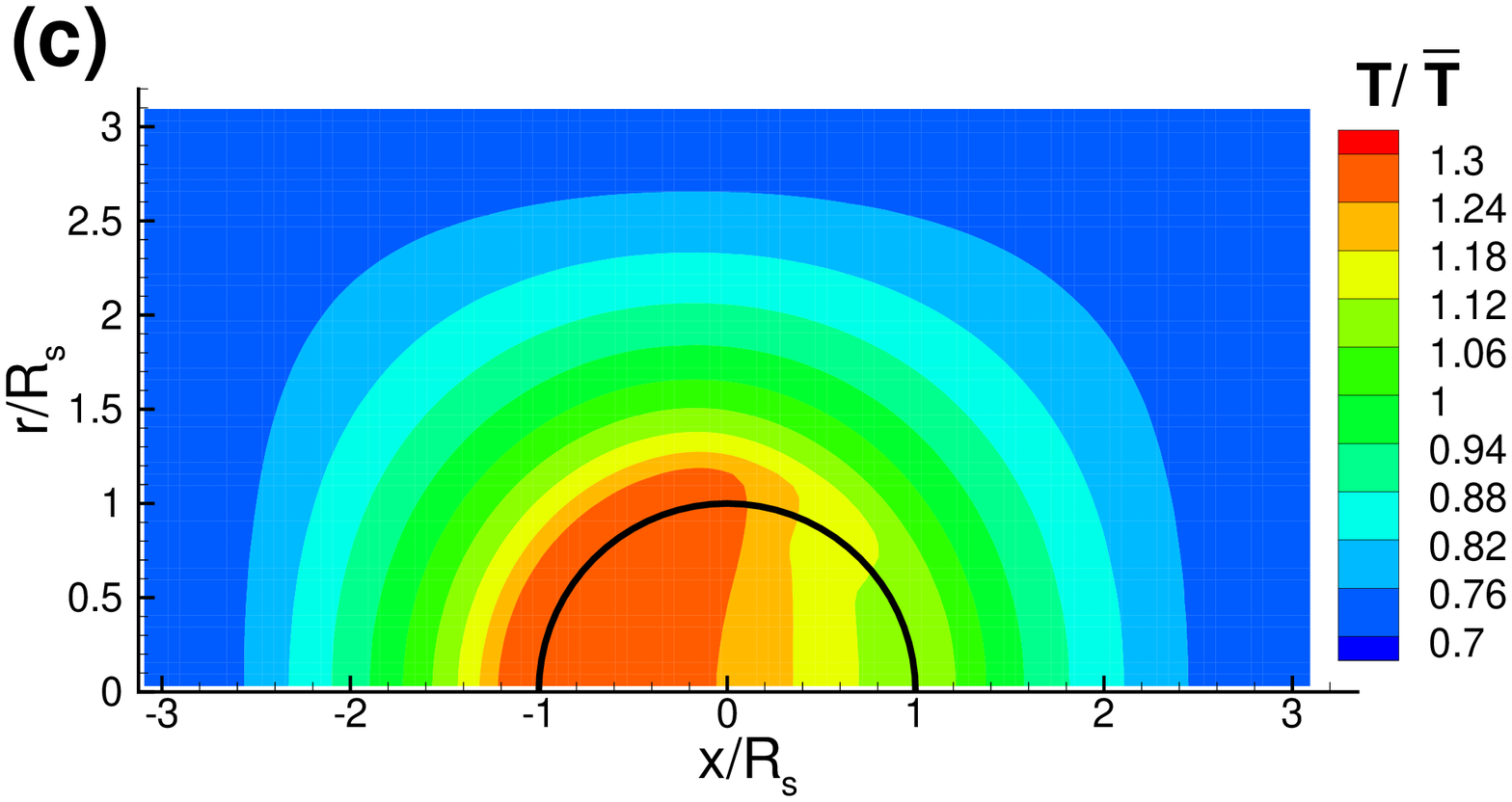}
\caption{Steady-state temperature profiles within the two-dimensional cross-section of a spherical Janus colloid
(black semi-circle) for (a) a differentially heated surface of the colloid with temperatures
$T_{hot}/\bar{T} = 1.3$ and $T_{cold}/\bar{T} = 0.7$ maintained on the hot and cold caps of the colloid,
respectively, (b) the colloid heated only at one hemispherical cap maintained at $T_{hot}/\bar{T} = 1.3$
and with a temperature control at the periodic boundaries in the $y-$direction maintained at
$T_{cold}/\bar{T} = 0.7$, and (c) the colloid heated only at one hemispherical cap maintained at $T_{hot}/\bar{T} = 1.3$
and with the temperature control at all periodic boundaries maintained at $T_{cold}/\bar{T} = 0.7$. 
Here, $\bar{T}=(T_{hot}+T_{cold})/2$ is the average temperature, which has been set to $\bar{T}=0.64$ in simulations.
These simulations correspond to the case of $a_F = 39 k_B\bar{T}/r_c$ and $a_{iA} = a_{iB} = 7.8 k_B\bar{T}/r_c$.}
\label{fig:temp}
\end{figure}

The temperature gradient across the colloid is maintained by setting an elevated temperature $T_{hot}$ for 
the immobile particles comprising cap $A$, while the particles comprising cap $B$ are kept at 
a lower temperature $T_{cold}$ throughout the course of the simulation, see Fig.~\ref{fig:temp}(a).
In experiments, however, one-half of the Janus colloid is heated by allowing the metal-capped half to
absorb heat from an incident laser field \cite{Jiang_AMP_2010,Buttinoni_ABM_2012}. The bulk fluid in this 
case remains at a colder temperature, and thus maintains a temperature gradient across the colloid diameter 
without the need to additionally cool the other half of the colloid. Therefore, we have also examined the case 
where we keep the particles of cap $A$ at an elevated temperature $T_{hot}$ and impose the temperature $T_{cold}$
for the fluid particles appearing within a narrow slice of $1.5 r_c$ at the two periodic boundaries parallel
to the asymmetry axis of the Janus colloid (Fig.~\ref{fig:temp}(b)). In this case, we let the temperature of
all other particles, including those of cap $B$, relax to an intermediate steady-state value.
Similarly, we have also compared these cases with the situation when all the periodic boundaries
are kept at the lower temperature $T_{cold}$ within a slice of $1.5 r_c$, while the particles of cap $A$ are maintained
at the higher temperature $T_{hot}$ (Fig.~\ref{fig:temp}(c)). We denote a maximal temperature difference across the 
colloid as $\Delta T = T_A-T_B$, which is equal to $T_{hot}-T_{cold}$ for the case where the temperature is controlled 
on both sides of the Janus colloid. For the other two cases, where the temperature is maintained only at one side of 
the colloid, $\Delta T = T_{hot}-T_B$ with $T_B$ measured directly from the simulation data. 
The differences in Janus-colloid behavior with 
respect to the different temperature-control strategies will be discussed. 

\subsection{Simulation setup and parameters}

In the simulations, a cubic box of size $L_x = L_y = L_z = 25r_c$ with periodic BCs is used.
We employ $m=1$ to define the unit of mass, $r_c=1$ to define the unit of length, and 
$\tau=r_c\sqrt{m/(k_B\bar{T})}=1.25$ is the unit of time, where $k_B=1$ and $\bar{T}=(T_{hot}+T_{cold})/2$ 
is the average temperature, which has been set to $\bar{T}=0.64$ in simulations. 
For the normalization of different colloid properties, we will also use the rotational 
diffusion coefficient $D_r=k_B\bar{T}/(8\pi\eta R_s^3)$, where $\eta$ is the fluid's dynamic viscosity at $\bar{T}$.    
Other parameters include the density of fluid particles $\rho_f =  3/r_c^3$, the 
conservative force coefficient $a_F=39 k_B\bar{T}/r_c$, the random force coefficient 
$\sigma_{ij} = 4.2 k_B\bar{T}\sqrt{\tau}/r_c$, the exponent $s=0.25$, the specific heat 
$C_v = 200 k_B$, and the nominal strength of interparticle heat conductivity $k_0 = 0.00125/\tau$.
The fluid viscosity for $a_F=39 k_B\bar{T}/r_c$ at $\bar{T}$ is equal to $\eta=2.74\sqrt{mk_B\bar{T}}/r_c^2$, 
and thus $D_r=2.27\times10^{-4}\sqrt{k_B\bar{T}/(mr_c^2)}$. We will also 
employ a fluid with $a_F=0$ whose viscosity is equal to $\eta=1.78\sqrt{mk_B\bar{T}}/r_c^2$ 
resulting in $D_r=3.5\times10^{-4}\sqrt{k_B\bar{T}/(mr_c^2)}$.  
The simulations were performed with discrete timesteps $\Delta t = 0.004\tau$ and data were collected 
after an initial relaxation time of $2 \times 10^5$ timesteps, in order to ensure a proper steady state. 
The relaxation to the steady state has been monitored by observing a stable time-independent temperature 
gradient and flow field. The steady-state averages were calculated by collecting data over another 
$8 \times 10^6$ timesteps.

\section{Results}

We study the dynamics of the model thermophoretic Janus colloid in the host fluid medium under various 
temperature conditions and surface interactions. The center of the Janus colloid is fixed at the origin 
of the reference frame as discussed in Sec.~\ref{sec:model} such that we can conveniently study 
the translational motion of the colloid by measuring the fluid velocity far from the colloid surface.
All our results are shown for steady-state conditions.

\subsection{Temperature dependence of self-propulsion}

Figure \ref{fig:temp} shows the temperature profiles around the colloid in the steady-state for 
the different temperature-control strategies described above. The temperature profiles are 
axisymmetric allowing us to average simulation data over the full azimuthal angle. Clearly, the temperature profiles are very different 
in the three cases. In Fig.~\ref{fig:temp}(a), a very strong temperature gradient develops at the interface 
between the two caps; this interfacial gradient is much smaller in Figs.~\ref{fig:temp}(b) and \ref{fig:temp}(c), which is both 
due to the smaller temperature difference between the two caps and the temperature variation on the
non-heated cap itself. The comparison of the profiles of Figs.~\ref{fig:temp}(b) and \ref{fig:temp}(c) shows minor
differences in the close vicinity of the colloid surface, but a much slower decay along the symmetry
axis in Fig.~\ref{fig:temp}(b).

\begin{figure*}
\centering
\includegraphics[scale=0.4]{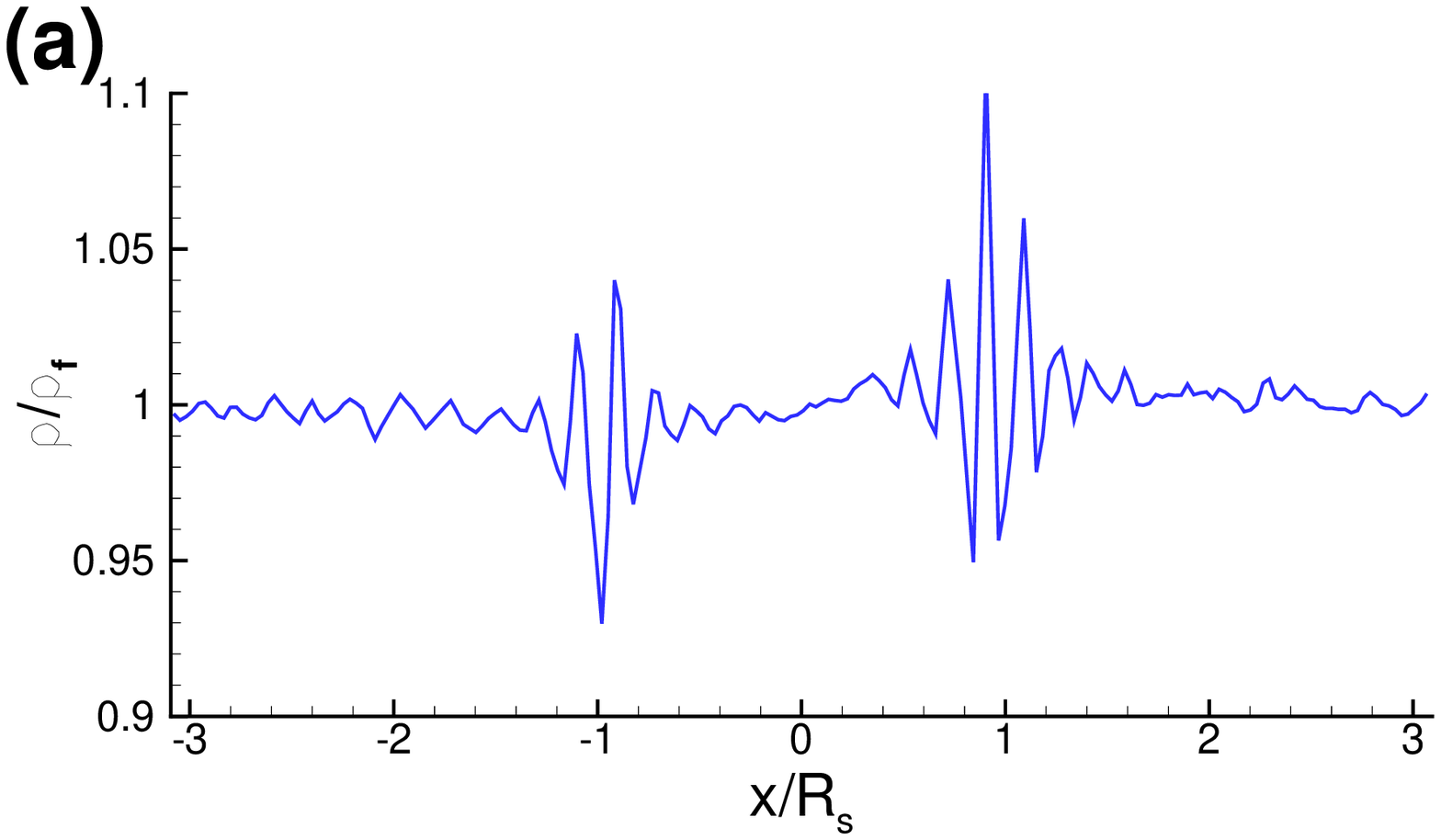}
\includegraphics[scale=0.4]{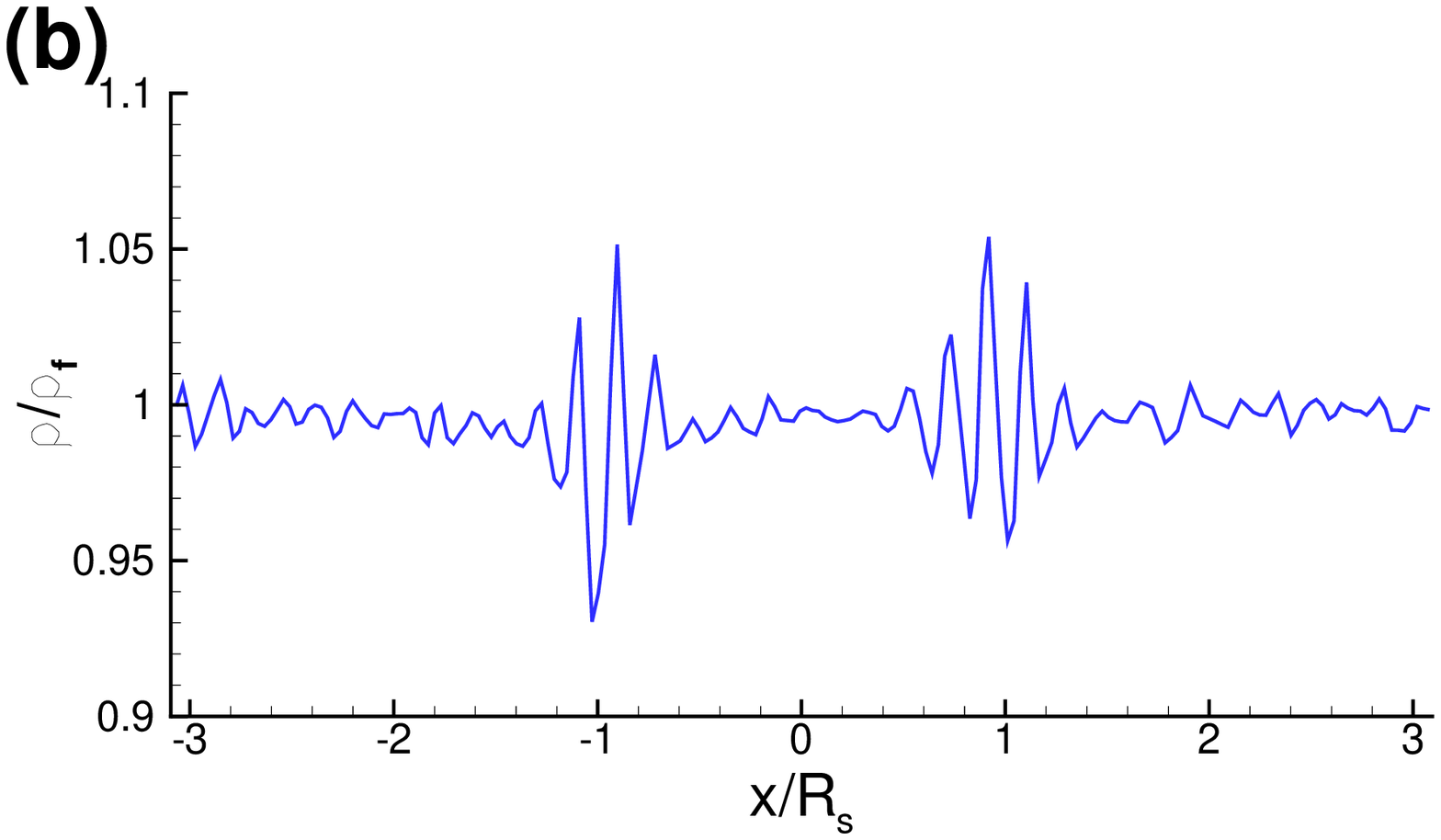}
\includegraphics[scale=0.4]{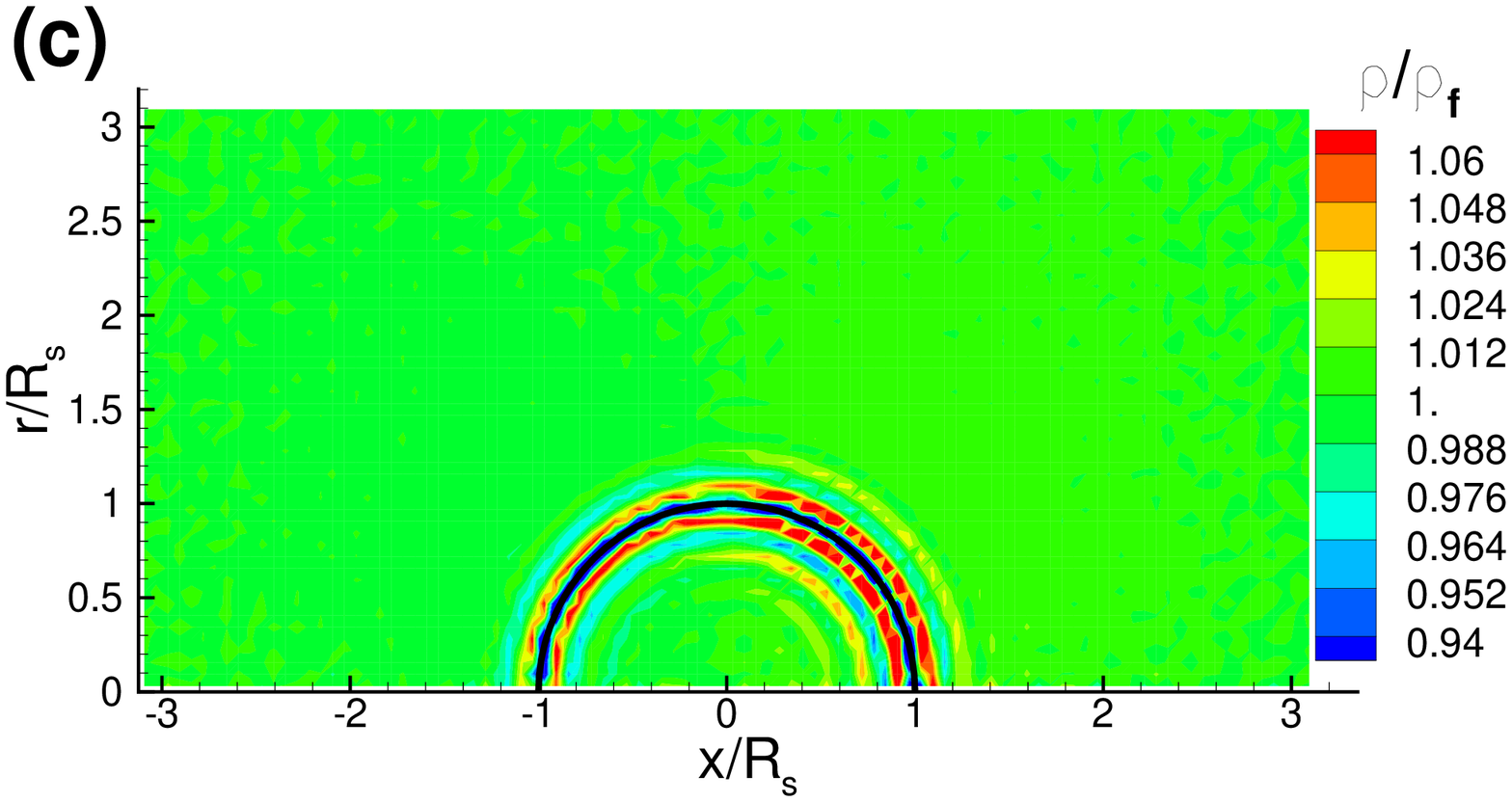}
\includegraphics[scale=0.4]{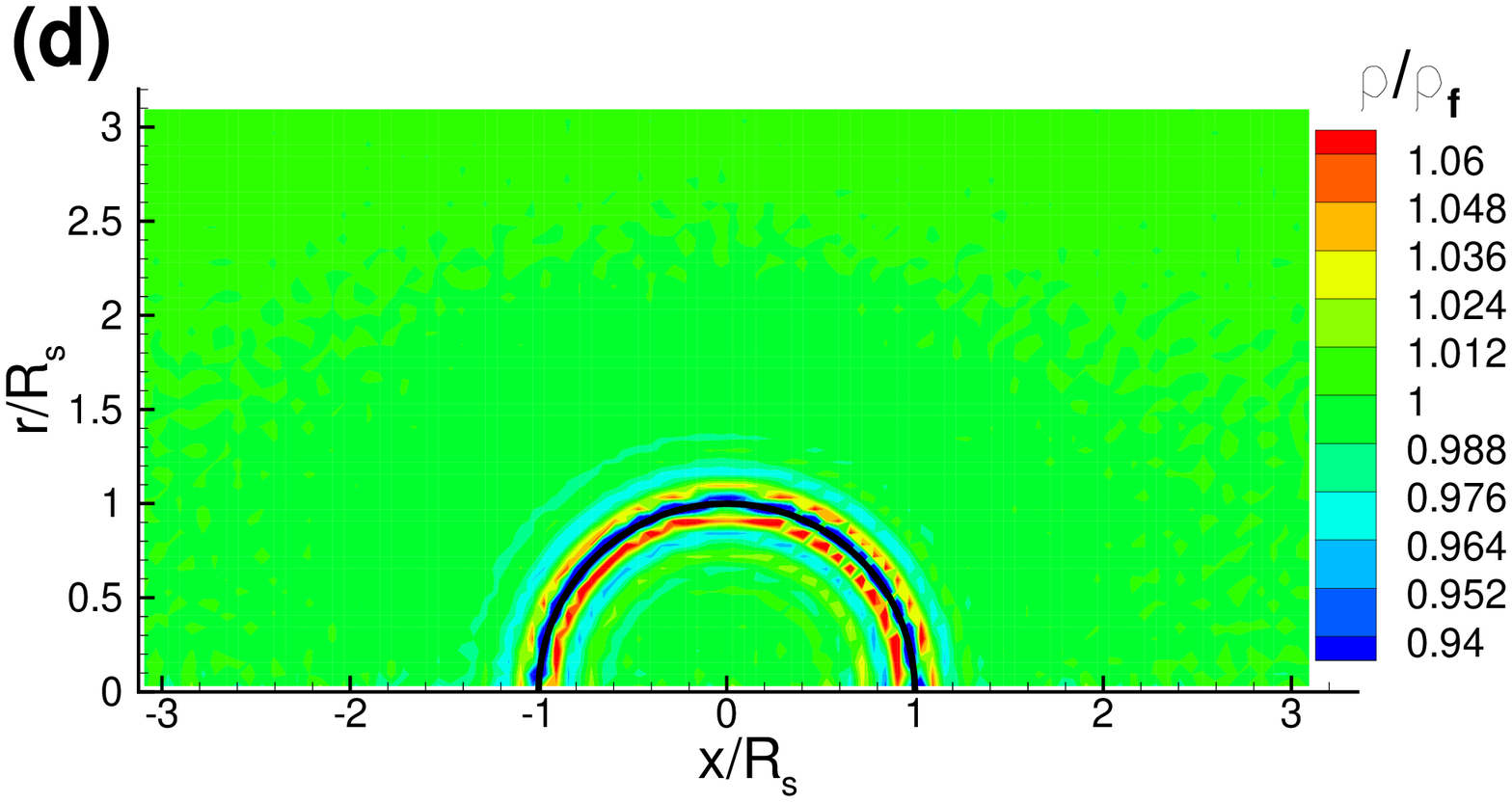}
\caption{Density ($\rho$) profile of the fluid around the Janus colloid in the steady-state.
The variation of the fluid density extracted along the $x-$axis at $r=0$ 
is shown for the Janus colloid with (a) a hot ($T_{hot}/\bar{T} = 1.3$) and a cold
($T_{cold}/\bar{T} = 0.7$) cap and (b) a hot cap ($T_{hot}/\bar{T} = 1.3$) and a temperature
fixed to be $T_{cold}/\bar{T} = 0.7$ at the periodic boundaries in the $y-$ direction. 
(c) and (d) show the corresponding density profiles in two dimensions averaged axisymmetrically 
over the full azimuthal angle. These simulations correspond to the case of $a_F = 39 k_B\bar{T}/r_c$, 
$a_{iA} = a_{iB} = 7.8 k_B\bar{T}/r_c$, and bounce-back reflection BCs.}
\label{fig:density}
\end{figure*}

In all three cases, we also measure the density profiles of the fluid near the colloid surface.
Figure \ref{fig:density} shows the fluid density cuts along the $x-$axis and the corresponding 
density profiles in two dimensions. We find a 
layering of the interacting fluid particles at the fluid-colloid surface. 
Such a layering is well known for hard-core particles near a hard wall \cite{Henderson_IBF_1984}, and is 
thus related to the repulsive interaction of fluid particles among themselves and with the 
colloid surface. These density modulations at the colloid wall
decay to the mean bulk density ($\rho_f = 3/r_c^3$) rapidly over a length scale of two to three times $r_c$.
However, the density patterns are similar for different temperature-control strategies. 
Also, density profiles do not show any appreciable differences for the various BCs, 
in particular the bounce-back and the specular reflection BCs at the fluid-colloid interface. 
The fluid density profiles in case of a hot cap ($T_{hot}/\bar{T} = 1.3$) and a temperature fixed to 
be $T_{cold}/\bar{T} = 0.7$ at the periodic boundaries in all directions are nearly identical to 
those in Figs. \ref{fig:density}(b) and \ref{fig:density}(d) for the temperature control only at the periodic boundaries 
in the $y$-direction.

\begin{figure}[h!]
\centering
\includegraphics[scale=0.4]{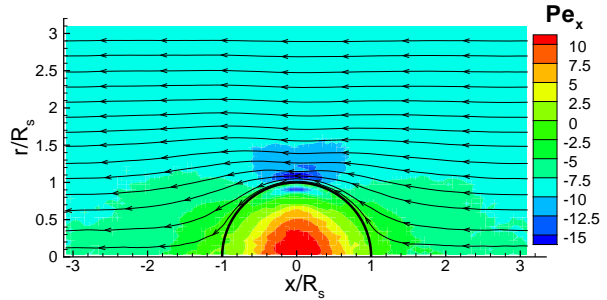}
\caption{Streamlines for fluid flow around the thermophoretic microswimmer with symmetric 
cap interaction ($a_{iA} = a_{iB} = 312 k_B\bar{T}/r_c$) in the host fluid ($a_F = 39 k_B\bar{T}/r_c$), 
and bounce-back BCs at the fluid-colloid interface (black semi-circle).
The color-code corresponds to the $x$-component of the fluid velocity ($v_x$) shown in terms 
of the local Peclet number $Pe_x=v_x/(2R_sD_r)$. This simulation corresponds to the case when the temperature is controlled at 
the colloid surface with a hot ($T_{hot}/\bar{T} = 1.3$) and a cold ($T_{cold}/\bar{T} = 0.7$) cap.}
\label{fig:flowStreamsVxBB}
\end{figure}

\begin{figure}
\vspace{-0.5cm}
\centering
\includegraphics[scale=0.31]{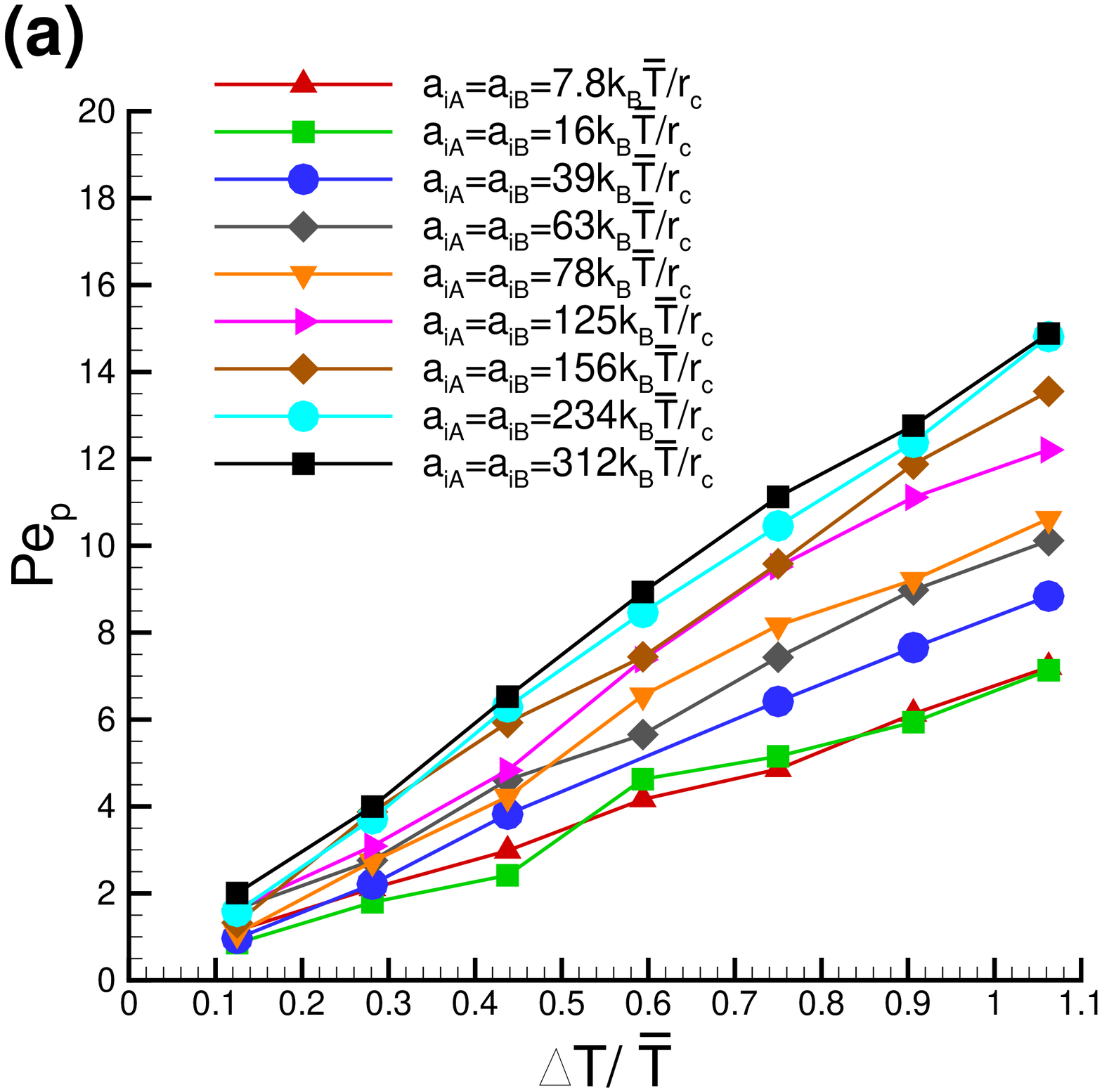}
\includegraphics[scale=0.31]{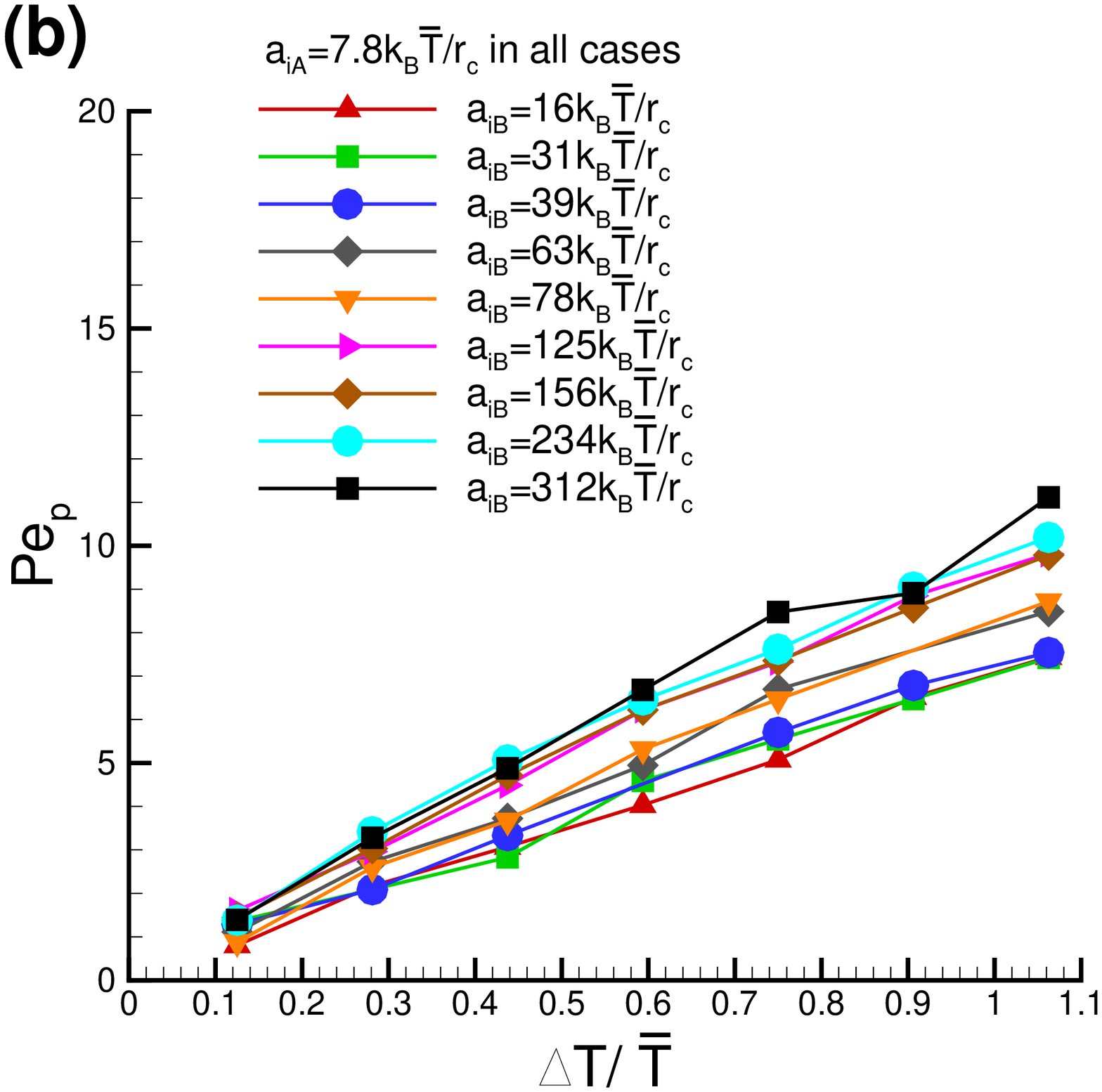}
\includegraphics[scale=0.31]{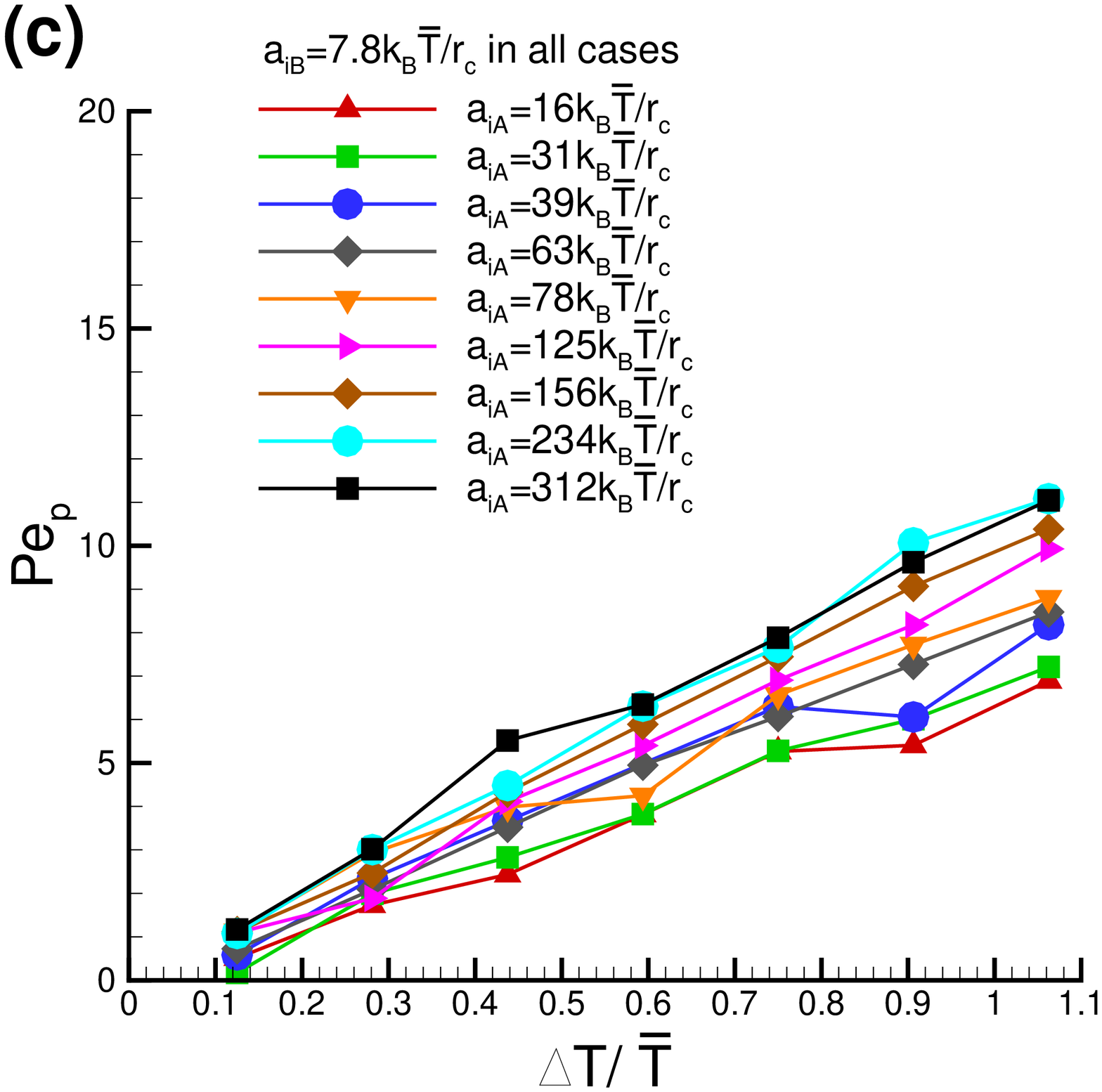}
\caption{Self-propulsion velocity $v_p$ presented in terms of the particle Peclet number 
$Pe_p=v_p/(2R_sD_r)$ as a function of the temperature difference,
$\Delta T = T_A - T_B$, for a fluid with interaction strength $a_F = 39 k_B\bar{T}/r_c$, when 
(a) $a_{iA} = a_{iB}$ (symmetric cap repulsion strengths), (b) $a_{iA} < a_F$ (an effectively solvophilic 
hot cap), and  (c) $a_{iB} < a_F$ (an effectively solvophilic cold cap). All data points are obtained for
the bounce-back BCs at the fluid-colloid interface. All simulations correspond to the case when the temperature is 
controlled directly at both caps of the colloid.}
\label{fig:VpDT}
\end{figure}

The non-zero temperature difference across the colloid diameter
results in the spontaneous generation of a far-field flow velocity in the fluid in the 
co-moving frame of the colloid (see Fig. \ref{fig:flowStreamsVxBB}). We measure this flow 
velocity by averaging over the velocities of all the DPD fluid particles
far away from the colloid with coordinates $|r| > 2 R_s$. In the laboratory 
frame, this flow velocity is the same in magnitude and
opposite in direction to the self-propulsion velocity $v_p$ of the colloid.
We find that $v_p$ increases linearly with increase in the
maximal temperature difference $\Delta T$ across the colloid at fixed average temperature 
$\bar{T}$, as shown in Fig.~\ref{fig:VpDT} 
in terms of a particle Peclet number $Pe_p=v_p/(2R_sD_r)$. $Pe_p$ can be also 
interpreted as a non-dimensional swimming velocity of the colloid.
Defining the corresponding proportionality constant as the thermophoretic
mobility, $\mu = v_p / \Delta T$, we can readily obtain $\mu$ from the slope.
The thermophoretic mobility is determined by the surface properties
of the Janus particle, the interactions within the fluid, and the average temperature 
$\bar{T}$, but obviously independent of $\Delta T$, and thus, it is a convenient quantity
to characterize a thermophoretic microswimmer. We can then use the thermophoretic mobility 
$\mu$ to study the dependence of self-propulsion on the surface properties of the colloid 
in relation to the host fluid. Subsequently, our simulation results will be presented mostly in terms 
of a non-dimensional mobility $\mu^*= Pe_p \bar{T}/\Delta T = \mu \bar{T}/(2R_sD_r)$.

\subsection{Dependence of thermophoretic mobility on fluid-colloid interactions}

\subsubsection{Effect of fluid-colloid repulsion.}

\begin{figure*}
\centering
\includegraphics[scale=0.33]{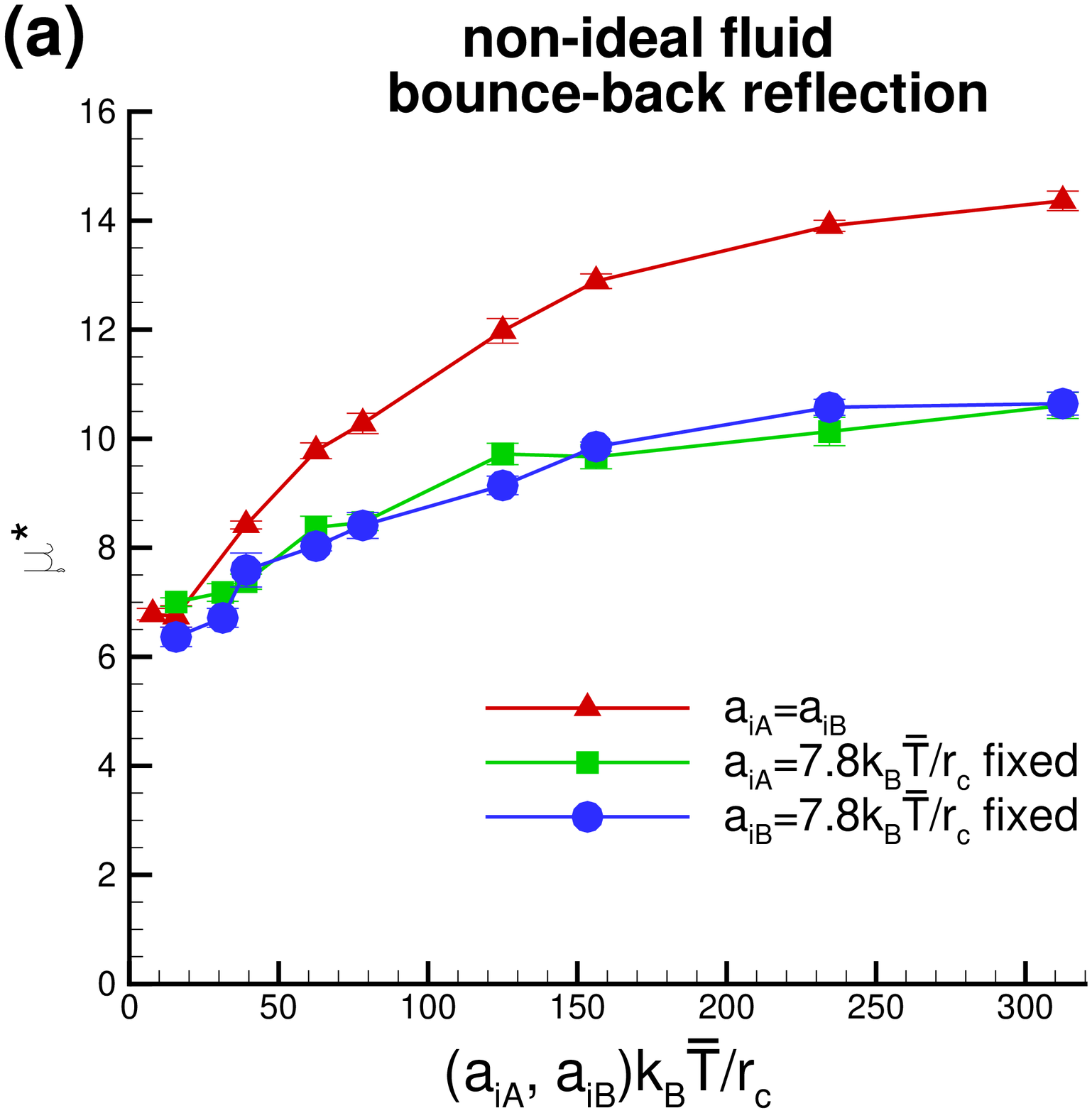}
\includegraphics[scale=0.33]{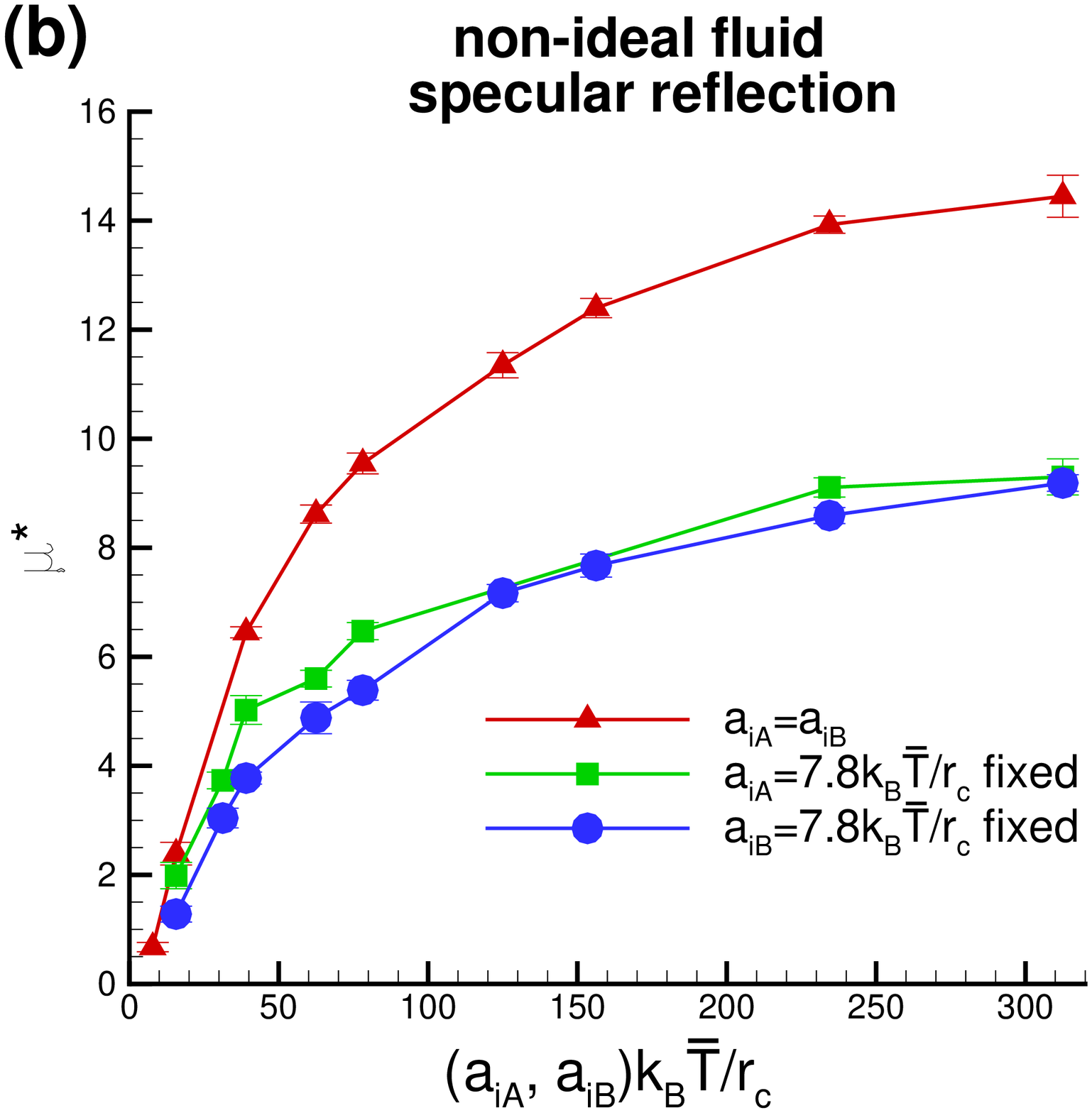}
\caption{Non-dimensional thermophoretic mobility $\mu^*$ as a function of fluid-cap repulsion 
strength with (a) bounce-back and (b) specular reflection BCs and for interacting fluid particles 
with $a_F = 39 k_B\bar{T}/r_c$. The hot and the cold caps ($A$ and $B$) are taken to be interacting 
with the fluid particles symmetrically with $a_{iA} = a_{iB}$ (red triangles),
and asymmetrically with $a_{iA} \le a_{iB}$ for $a_{iA} = a_F/5$ and varying $a_{iB}$ (solvophilic hot cap, green squares) 
and with $a_{iA} \ge a_{iB}$ for $a_{iB} = a_F/5$ and varying $a_{iA}$ (solvophilic cold cap, blue circles).
All simulations correspond to the case when the temperature is controlled at the colloid surface with 
a hot ($T_{hot}/\bar{T} = 1.3$) and a cold ($T_{cold}/\bar{T} = 0.7$) cap.}
\label{fig:MuCapRepRealFluid}
\end{figure*}

\hspace{0.2cm}
The surface properties of the Janus colloid can be tuned to manipulate and control 
its thermophoretic mobility. We first consider the case of ``symmetric caps'', when 
--- apart from the different temperature conditions at the two caps ($T_A > T_B$) --- 
the surface interaction of both caps $A$ and $B$ with the fluid particles $i$ are
identical, i.e., $a_{iA} = a_{iB} \equiv a_{cap}$. The dependence of the self-propulsion velocity 
on $\Delta T$ is shown in Fig. \ref{fig:VpDT}(a) for a wide range of cap interactions, with 
$0 < a_{cap}/a_{F} \le 8$. These results indicate that the thermophoretic mobility depends highly 
non-linearly on the fluid-cap repulsion strength, $a_{cap}$, as shown in 
Fig.~\ref{fig:MuCapRepRealFluid} for the case of $\Delta T/\bar{T}=0.6$. 
The mobility increases with increasing cap repulsion strength, and levels off to some saturation 
value at very high repulsion strength.

Figure \ref{fig:MuCapRepRealFluid} reveals an interesting dependence of $\mu^*$ on the slip or stick BCs. 
Within the numerical accuracy, we find that the saturation value at high repulsion strength 
does not depend on whether bounce-back or specular-reflection collisions are employed at 
the colloid surface. This can easily be understood by considering the fact that at high repulsion 
strengths hardly any fluid particles can reach the colloid surface anymore, so that the type of 
surface reflection becomes irrelevant. However, for low cap repulsion strengths, the
stick BCs generate larger fluid flow (and thus, propulsion velocity) than the slip BCs for 
a given temperature difference and cap repulsion strength. In fact, in the absence of cap repulsion, 
the thermophoretic mobility disappears in the case of slip BCs.

Figure \ref{fig:VpDT} also shows the velocity-temperature graphs for asymmetric cap 
repulsion strengths. Figure \ref{fig:VpDT}(b) concerns the case when the hot cap
$A$ interacts with a lower interaction coefficient with the fluid particles compared to
the fluid-fluid interaction ($a_{iA} < a_F$), to mimic a solvophilic cap interaction.
This interaction is kept fixed and the interaction of the cold cap $B$ with the
fluid is increased from solvophilic ($a_{iB} < a_F$) to solvophobic ($a_{iB} > a_F$) strengths.
In Fig. \ref{fig:VpDT}(c), the cold cap $B$ is instead maintained at a solvophilic interaction 
strength ($a_{iB} < a_F$), and the interaction $a_{iA}$ on the hot cap $A$ is varied.
These two cases are not identical, because the $A$-cap is always the hot and the $B$-cap is 
always the cold side. The velocity response to the temperature difference remains linear for 
all cases. The non-dimensional thermophoretic mobility $\mu^*$ extracted from these data for the case of $\Delta T/\bar{T}=0.6$ is
shown in Fig. \ref{fig:MuCapRepRealFluid}. As for the symmetric case, $\mu^*$ is found to increase 
with increasing repulsion strength $a_{iA}$ or $a_{iB}$. Figs.~\ref{fig:MuCapRepRealFluid}(a) and \ref{fig:MuCapRepRealFluid}(b)
also provide the comparison of mobilities for stick and slip BCs, which show similar trends as for the 
symmetric case, with an increase in $\mu^*$ and its final saturation as the cap repulsion strength is 
elevated.

The thermophoretic mobility obtained from these two very different asymmetric fluid-cap interaction 
cases remains essentially identical (within our numerical accuracy), although it is much lower 
than the corresponding mobility for symmetric fluid-cap interactions (where the larger value of the two repulsion 
strengths in the asymmetric case is the same as $a_{cap}$ in the symmetric case). 
In particular, the saturation value of $\mu^*$ in the asymmetric case is considerably smaller
than that for the symmetric case. Furthermore, Fig. \ref{fig:MuCapRepRealFluid} shows that 
the increase of the mobility due to cap repulsion is about twice as large for the symmetric than 
for the (highly) asymmetric cases. Together, this indicates that the thermophoretic mobility 
can be understood as a superposition of the mobilities generated by the two caps independently. 
This indicates that the temperature gradient between the cap and the fluid, rather than that at 
the interface between the two caps, is responsible for the propulsion. Figure \ref{fig:MuCapRepRealFluid} 
also shows that the insensitivity of the thermophoretic mobility to the exchange of repulsion strengths 
is also not affected by the type of BCs at the colloid-fluid interface. However, the slip BCs always generate 
lower mobilities compared to stick boundary BCs, in particular for lower values of fluid-cap interaction 
strengths.

\subsubsection{Effect of temperature control.}

\begin{figure}
\centering
\includegraphics[scale=0.33]{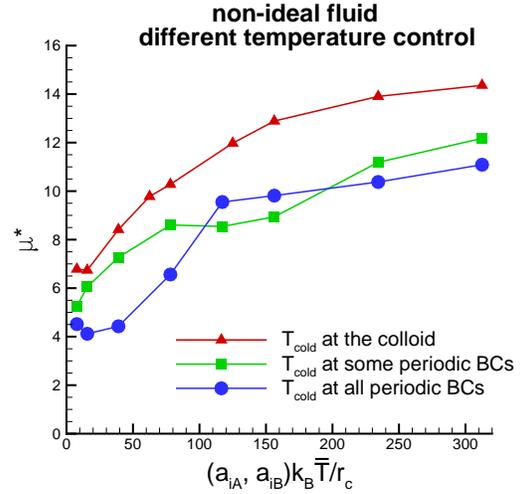}
\caption{Non-dimensional thermophoretic mobility $\mu^*$ as a function of fluid-cap 
repulsion strength with bounce-back BCs, for interacting fluid particles with 
$a_F = 39 k_B\bar{T}/r_c$ and with a temperature gradient maintained between one of the caps 
($A$) kept at an elevated temperature ($T_{hot}=1.3\bar{T}$) and the periodic boundaries maintained at 
a fixed lower temperature ($T_{cold}=0.7\bar{T}$). The green squares represent the thermophoretic mobilities 
when two opposite periodic boundaries parallel to the symmetry axis of the Janus colloid are kept 
at the temperature $T_{cold}$, and the blue circles correspond to the case when the temperature 
at all six periodic boundaries is maintained at $T_{cold}$. For comparison, red triangles represent the 
case where the temperature $T_{cold}$ is maintained directly at cap B. The cap repulsions are taken to be 
symmetric ($a_{iA} = a_{iB}$) for all cases.}
\label{fig:MuCapRepRealFluidHotOnly}
\end{figure}

\hspace{0.2cm}
Figure \ref{fig:MuCapRepRealFluid} reflects the dependence of the thermophoretic mobility when 
both sides of the Janus colloid are maintained at fixed temperatures $T_A=T_{hot}=1.3\bar{T}$ and $T_B=T_{cold}=0.7\bar{T}$ 
(illustrated in Fig.~\ref{fig:temp}(a)). We now consider the case where the $A$-cap is held at a 
fixed temperature $T_{hot}$, and the fluid far away from the colloid (and near the periodic boundaries 
of the simulation box) is held at a lower value $T_{cold}$, while the temperature of the other cap 
($B$) is free to adjust, as illustrated in Figs.~\ref{fig:temp}(b) and \ref{fig:temp}(c). 
In this case, $T_B$ is measured directly from the simulation data. The temperature distribution is nearly 
independent of the fluid-colloid interactions and the type of BCs, and the measured values of temperature 
at the cold side are $T_B=1.11\bar{T}$ when the two periodic boundaries parallel to the symmetry axis 
of the Janus colloid are kept at the temperature $T_{cold}$ and $T_B=1.09\bar{T}$ when all periodic 
boundaries are maintained at $T_{cold}$. The results for the non-dimensional 
thermophoretic mobility displayed in Fig. \ref{fig:MuCapRepRealFluidHotOnly} 
show that the qualitative trends remain unaltered. However, $\mu^*$ is slightly lower
than for the case of fixed temperature $T_B=T_{cold}$. This is due to the fact that temperature gradients 
at the interface between the cold and hot sides of the colloid are stronger for the case of 
the temperature control $T_B=T_{cold}$ directly at the colloid.   

\subsubsection{Comparison with previous simulations.}   

\hspace{0.2cm}
Over the entire range of temperature and interaction potentials considered, we do not observe
any reversal of the propulsion direction. The colloid is always propelled in the direction of
the cold side. Recent simulations \cite{Yang_TNS_2011,deBuyl_PSP_2013,Yang_HSS_2014} using 
another mesoscale hydrodynamics simulation technique, the multiparticle
collision dynamics (MPC) method \cite{Malevanets_MSM_1999,Gompper_APS_2009}, have shown that the swimming 
direction of a thermophoretic colloid can be affected by the solvent-colloid interaction potential. 
In these studies \cite{Yang_TNS_2011,Yang_HSS_2014} only a symmetric case was considered, and 
an attractive potential between solvent particles and the thermophoretic colloid has led to swimming 
in the direction of the cold side, while a repulsive potential has triggered the colloid to move 
in the direction of the hot side. In our simulations, the conservative potential is purely repulsive; 
however, for the conditions $a_{iA}<a_F$ and $a_{iB}<a_F$, fluid particles should be effectively attracted
to the colloid surface. The current simulation results do not lead to a similar behavior of the Janus 
colloid where the swimming direction can be interchanged. It is important to note that the two fluid models 
are not equivalent. The main difference between eDPD and MPC 
fluids is that the MPC fluid has the equation of state of an ideal gas, while the eDPD fluid is much 
less compressible due to the presence of conservative interactions if $a_F > 0$. 
Furthermore, the heat capacity and thermal conductivity in these models are not equivalent, which will 
be discussed further in text.    

\begin{figure*}
\centering
\includegraphics[scale=0.33]{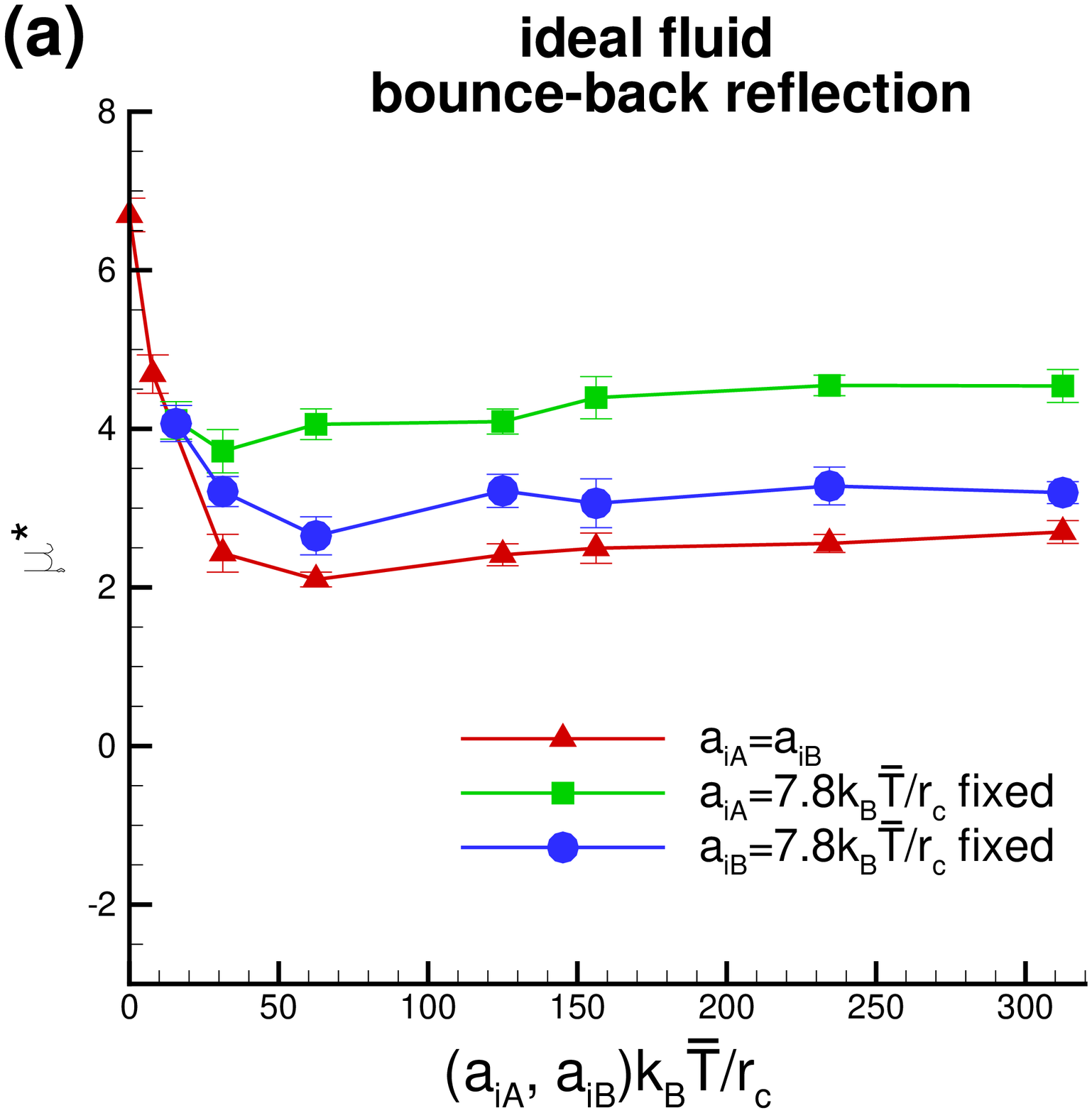}
\includegraphics[scale=0.33]{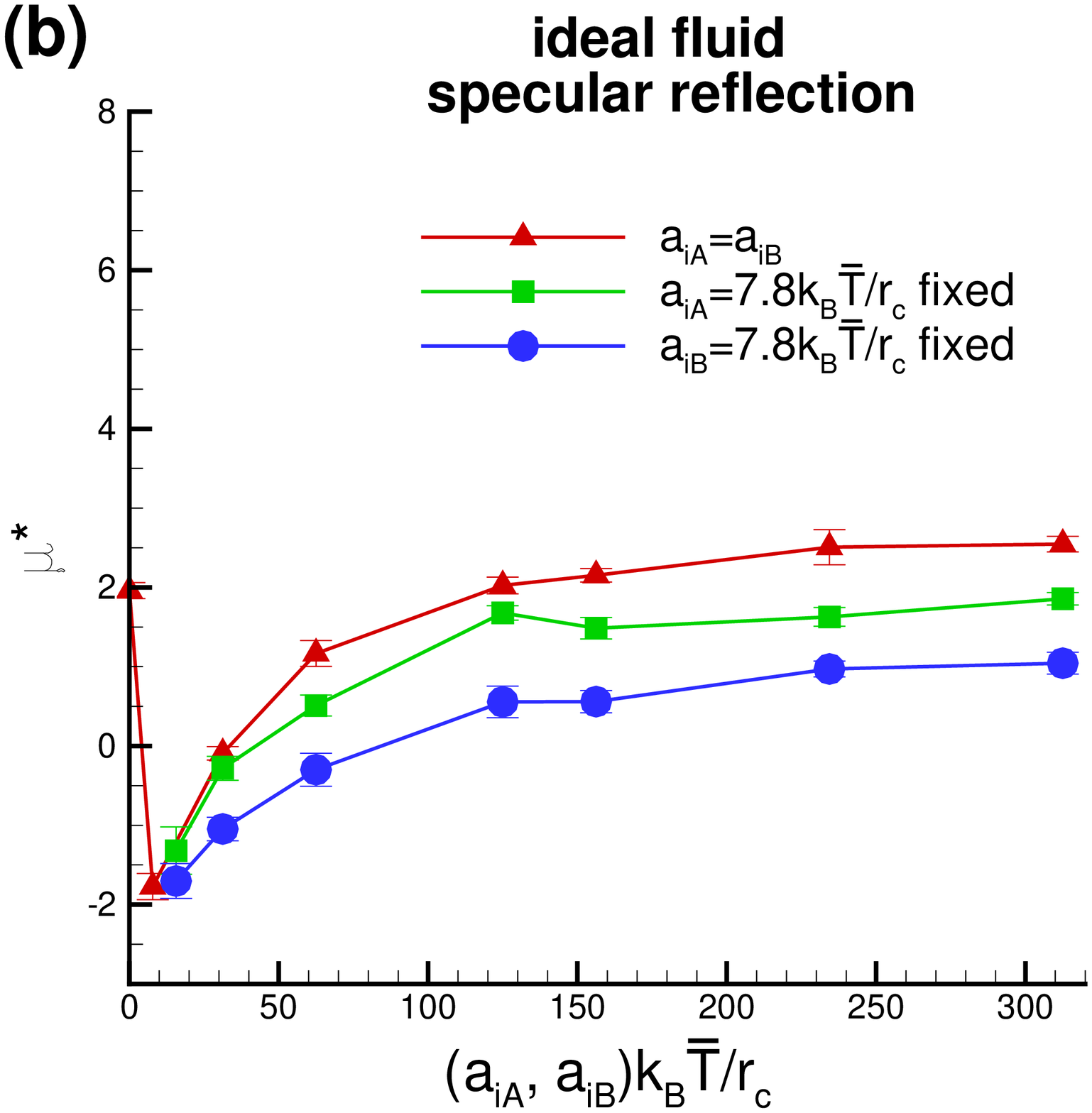}
\caption{Non-dimensional thermophoretic mobility ($\mu^*$) as a function of the fluid-cap repulsion 
strength with (a) bounce-back and (b) specular reflection BCs for ideal-gas fluid with $a_F = 0$.
The hot and the cold caps ($A$ and $B$) assume symmetric interactions $a_{iA} = a_{iB}$ with the fluid
(red triangles) as well as asymmetric interactions with a fixed $a_{iA}=7.8k_B\bar{T}/r_c$ and 
$a_{iB}$ varied (green squares) and with a fixed $a_{iB}=7.8k_B\bar{T}/r_c$ and $a_{iA}$ varied
(blue circles). All simulations correspond to the case when the temperature is controlled at 
the colloid surface with a hot ($T_{hot}/\bar{T} = 1.3$) and a cold ($T_{cold}/\bar{T} = 0.7$) cap.}
\label{fig:MuCapRepIdealFluid}
\end{figure*}

\subsection{Thermophoretic propulsion in ``ideal'' fluids}

The coefficient $a_F$ of the conservative interaction between fluid particles 
plays an important role in determining the compressibility of the fluid, but also
affects the diffusion coefficient and the fluid viscosity. Here, decreasing $a_F$ increases
the fluid compressibility. Further, we investigate the extreme case when $a_F = 0$, i.e., 
the case obtained when the fluid is most compressible. In this case, the equation of state is 
that of an ideal gas, similar to the MPC fluid model in Refs. \cite{Yang_TNS_2011,Yang_HSS_2014}. 
The thermophoretic mobility of the Janus colloid is then again measured in this host fluid medium for 
both bounce-back and specular reflection BCs, and its dependence on the fluid-colloid interaction 
strength for both symmetric and asymmetric cap cases is determined. 

In the case of bounce-back BCs at the colloid-fluid interface, the thermophoretic mobilities 
as a function of the fluid-cap interaction strength are shown in Fig. \ref{fig:MuCapRepIdealFluid}(a).
We find an initial decrease of the mobility with increasing cap-fluid interaction strength, 
followed by a subsequent weak recovery, and finally saturation with further increase of the
interaction strength, for both symmetric and asymmetric fluid-cap interactions.
This dependence of the thermophoretic mobility is qualitatively different from the behavior of 
a Janus colloid in a non-ideal fluid (with $a_F > 0$). Also, the propulsion velocity in the ideal 
fluid for the symmetric cap case remains lower than that for asymmetric cap, again in contrast to 
the non-ideal fluid case. Our results show that in addition to an enhancement of mobility 
with cap asymmetry, the magnitude can be further controlled by interchanging the temperature 
of the caps, unlike the non-ideal fluid scenario.

For specular-reflection BCs in Fig. \ref{fig:MuCapRepIdealFluid}(b), we obtain a reversal 
of propulsion direction of the colloid in the ideal fluid by tuning the fluid-cap interaction.
In the low fluid-cap interaction regime, the computed mobility is negative, which
is equivalent to  swimming in the direction towards the colloid's hot side. 
With increasing fluid-cap interaction strength, the corresponding mobility changes sign 
and becomes positive, i.e. the propulsion direction is inverted and the colloid swims toward 
its cold side. Over the entire range of the fluid-cap interactions studied in 
Fig. \ref{fig:MuCapRepIdealFluid}(b), the mobility increased gradually with the interaction 
strength before leveling off at very large fluid-cap interactions.

A comparison of Figs. \ref{fig:MuCapRepIdealFluid}(a) and \ref{fig:MuCapRepIdealFluid}(b) reveals that the 
trends of the mobilities over the entire range of cap repulsions for specular reflections 
and bounce-back collisions are quite different. In particular, the saturation value of 
the mobility for the colloid with symmetric caps remains higher than that with asymmetric 
caps for specular reflections unlike the bounce-back situation. Furthermore, interchanging the 
temperatures of the asymmetric caps leads to a change in the propulsion velocity, quite unlike 
the situation of non-ideal fluids.

In comparison to the MPC simulations in Refs. \cite{Yang_TNS_2011,Yang_HSS_2014}, we also
observe a reversal in the swimming direction by tuning the fluid-cap interaction strength. 
However, the direction reversal occurs only for the case of specular reflections (i.e., slip BCs)
and relatively small fluid-cap interaction strength. In these simulations, the fluid-cap 
interaction can be considered purely repulsive, since $a_F=0$ of the suspending fluid. Thus, 
in case of bounce-back reflections the colloid always swims toward the cold side similar 
to the results for non-ideal fluid case, while in case of slip BCs, the colloid moves toward 
the cold side for strong fluid-cap repulsion and toward the hot side for weak fluid-cap repulsive
interactions. This trend seems to be rather opposite to that in Refs. \cite{Yang_TNS_2011,Yang_HSS_2014},
where fluid-colloid repulsive interactions lead to a swimming direction toward the hot side, 
while fluid-colloid attractive interactions result in swimming toward the cold side. 
In addition, the swimming direction of a Janus thermophoretic colloid has not been affected by 
the type of BCs (i.e., slip or no-slip) in the MPC simulations \cite{Yang_TNS_2011,Yang_HSS_2014}.         
   
\begin{figure}
\centering
\includegraphics[scale=0.4]{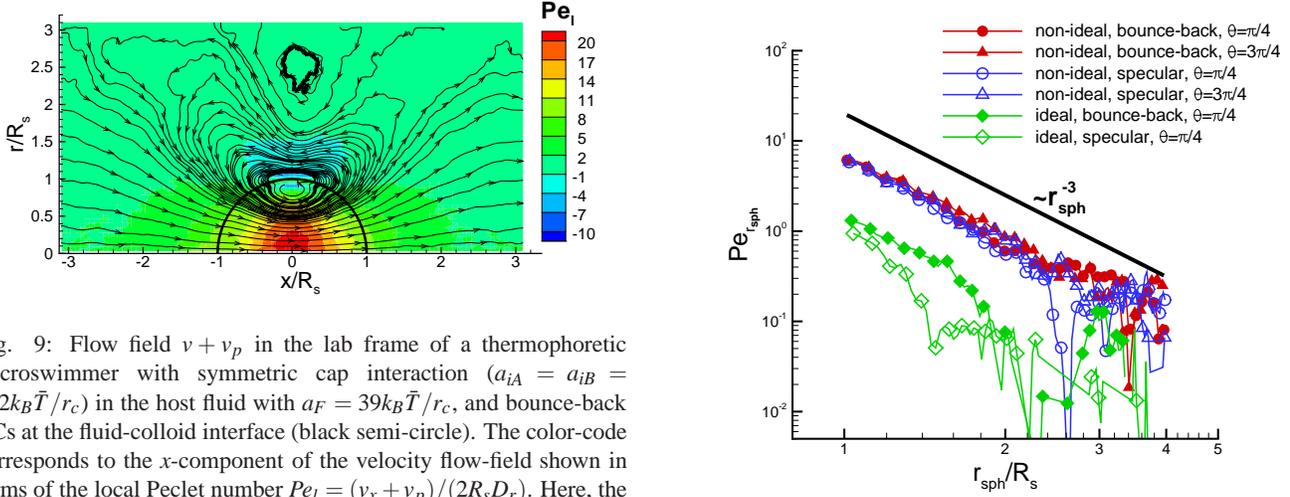}
\caption{Flow field $v + v_p$ in the lab frame of a thermophoretic microswimmer with 
symmetric cap interaction ($a_{iA} = a_{iB} = 312 k_B\bar{T}/r_c$) in the host fluid with 
$a_F = 39k_B\bar{T}/r_c$, and bounce-back BCs at the fluid-colloid interface (black semi-circle).
The color-code corresponds to the $x$-component of the velocity flow-field shown in terms of the 
local Peclet number $Pe_l=(v_x+v_p)/(2R_sD_r)$. Here, the temperature $T_{cold}$ is controlled 
directly at the colloid surface.}
\label{fig:flowFieldVBB}
\end{figure}

\subsection{Flow field around thermophoretic Janus colloids}

As the colloid swims, the flow field generated in the fluid can be found from our
simulations. Figure \ref{fig:flowFieldVBB} shows the flow field in the lab frame,
obtained by adding the velocity of self-propulsion $v_p$ to the fluid velocity $v$,
since its magnitude has an opposite sign to the far-field fluid velocity 
(see Fig. \ref{fig:flowStreamsVxBB} for example). The flow field indicates
that the source-dipole contribution dominates, which is consistent with the 
conclusions from theory \cite{Bickel_FPV_2013} and simulations \cite{Yang_HSS_2014,Yang_TPF_2013}
for a thermophoretic colloidal swimmer. The flow field in Fig. \ref{fig:flowFieldVBB}
is very similar to the theoretical predictions for a thin-cap limit \cite{Bickel_FPV_2013}
(i.e., the thermal conductivity of the cap does not play a role),
supporting the validity of the simulation results. The corresponding flow-field in MPC 
simulations \cite{Yang_HSS_2014,Yang_TPF_2013} is very close to a source-dipole
approximation in the theory \cite{Bickel_FPV_2013}.    
 
\begin{figure}
\centering
\includegraphics[scale=0.33]{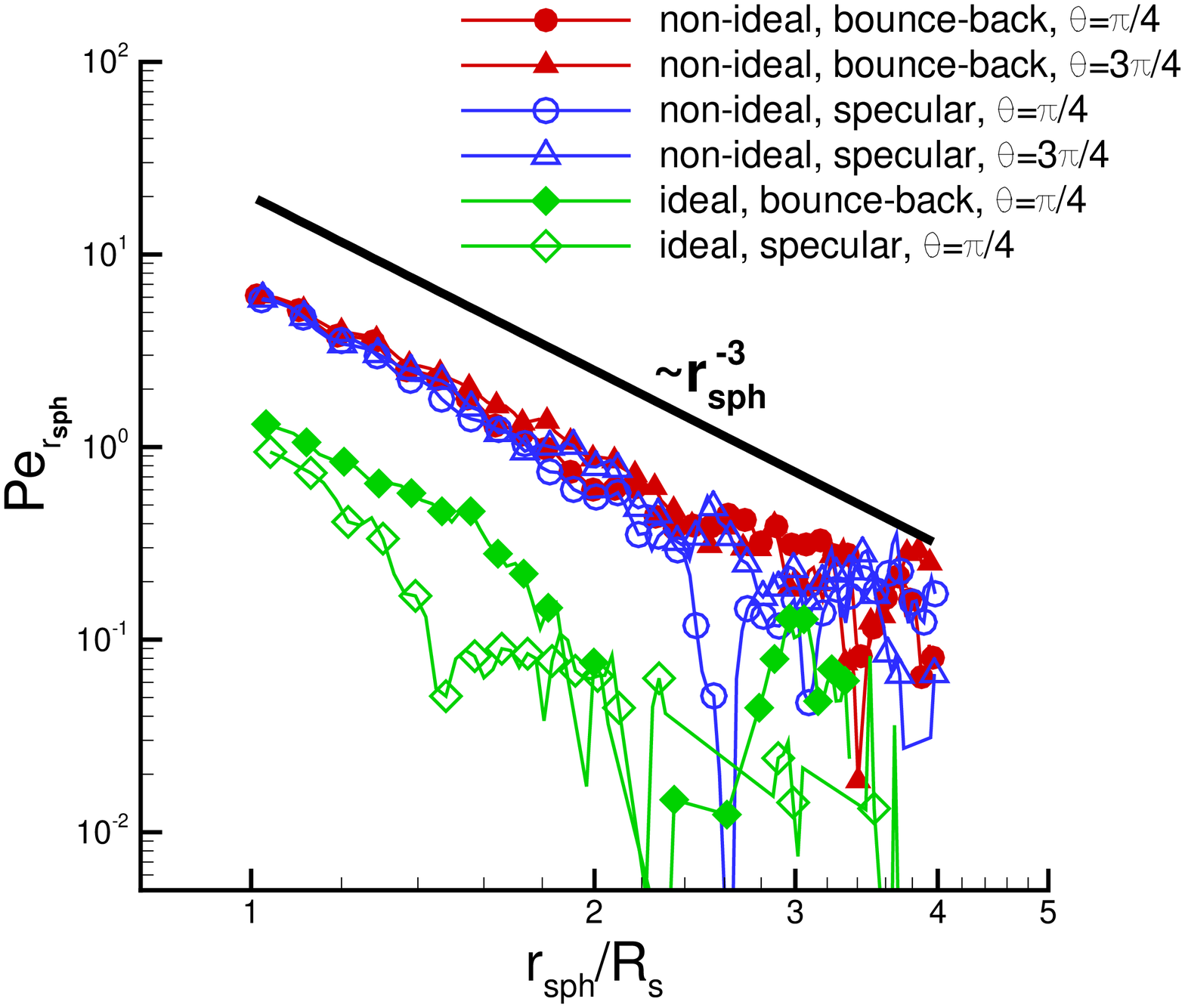}
\caption{Radial velocity $v_{r_{sph}}$ of the flow field $v + v_p$ in spherical coordinates 
presented in terms of the local Peclet number $Pe_{r_{sph}}=v_{r_{sph}}/(2R_sD_r)$ as a function 
of distance $r_{sph}$ from the center of the colloid, obtained along fixed polar angles, 
$\theta$, with respect to the symmetry axis of the Janus colloid.  The filled and 
the open symbols are for bounce-back and specular reflection BCs, respectively. 
Circles and triangles correspond to $\theta = \pi/4$ and $\theta = 3\pi/4$ cases, 
respectively, for a DPD fluid with $a_F = 39k_B\bar{T}/r_c$, and the diamonds correspond 
to an ideal gas fluid with $a_F = 0$ and $\theta = \pi/4$.  The thick solid line shows 
the power-law behavior $\sim r_{sph}^{-3}$. The simulations correspond to the 
case of $a_{iA} = a_{iB} = 312 k_B\bar{T}/r_c$ with the temperature $T_{cold}$ controlled 
directly at the colloid surface.}
\label{fig:VrPowerLaw}
\end{figure}

\begin{figure}
\centering
\includegraphics[scale=0.33]{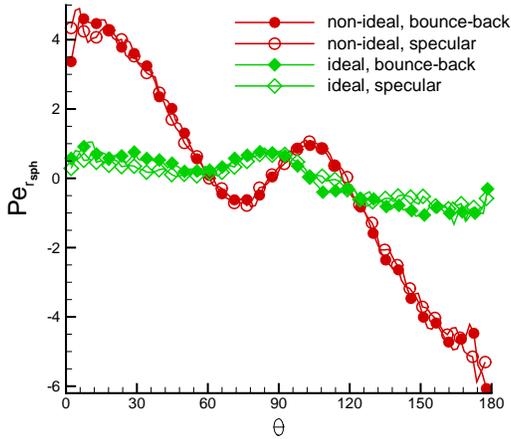}
\caption{Radial velocity $v_{r_{sph}}$ of the flow field $v + v_p$ in spherical coordinates    
presented in terms of the local Peclet number $Pe_{r_{sph}}=v_{r_{sph}}/(2R_sD_r)$ as a function of the angle 
($\theta$) with respect to the symmetry axis of the Janus colloid, measured at a distance $r_{sph} = 1.5 R_s$ 
from the center of the colloid. The diamonds correspond to an ideal-gas fluid with $a_F = 0$,
and the circles correspond to a DPD fluid with $a_F = 39k_B\bar{T}/r_c$. The filled and the open symbols
correspond to the cases with bounce-back and specular reflection BCs, respectively. Here, the simulations 
are performed using $a_{iA} = a_{iB} = 312 k_B\bar{T}/r_c$ and the temperature $T_{cold}$ is controlled   
directly at the colloid surface.}
\label{fig:VrTheta}
\end{figure}

Finally, we determine the radial component $v_{r_{sph}}$ of the velocity field $v + v_p$ in spherical coordinates 
as a function of the distance from the center of the colloid, see Fig. \ref{fig:VrPowerLaw}. 
The radial component of the velocity is measured at a particular angle, $\theta$, relative 
to the symmetry axis of the colloid. Within the statistical accuracy of our simulations, we 
observe a power-law dependence with $r_{sph}$, which is consistent with an inverse cubic power, 
$v_{r_{sph}} \sim 1/r_{sph}^3$, for both an ideal-gas fluid with $a_F = 0$, and a DPD fluid with non-zero $a_F$. 
This behavior is unaffected by the different BCs used, and is in good agreement with the theoretical
predictions \cite{Bickel_FPV_2013} and MPC simulations \cite{Yang_HSS_2014,Yang_TPF_2013}.
We also compute the radial velocity as a function of the angle $\theta$ at a distance $r_{sph} = 1.5 R_s$
from the center. The results are shown in Fig. \ref{fig:VrTheta} for different fluid interaction 
strengths and different BCs. We find an asymmetric radial velocity distribution around the colloid 
from the cold cap pole ($\theta = 0^\circ$) to the hot cap pole ($\theta = 180^\circ$).
This dependence is in qualitative agreement with the theoretical predictions of Ref. \cite{Bickel_FPV_2013}.

\section{Discussion and conclusions}

The swimming velocity of a thermophoretic Janus colloid strongly depends on different fluid-colloid 
interactions and BCs. The choice of BCs (i.e., slip or no-slip) affects viscous friction exerted on 
the colloid by the fluid, while different strengths of fluid-colloid repulsive interactions alter both, 
the heat exchange between the fluid and the colloid surface and the near-wall density fluctuations illustrated 
in Fig. \ref{fig:density}. Thus, the repulsive strengths $a_{iA}$ and $a_{iB}$ also have an effect on 
the viscous friction for the case of bounce-back reflections. Perhaps, the simplest case is that with 
specular reflections in Fig. \ref{fig:MuCapRepRealFluid}(b), since viscous friction between the colloid 
and the fluid can be neglected. In this case, the colloid mobility increases with increasing interaction 
strength, or equivalently when fluid particles are pushed further away from the surface of the swimmer. We 
expect that a larger distance between fluid particles and the colloid surface should lead to a reduction 
of heat exchange between the colloid and the fluid. In Fig. \ref{fig:MuCapRepRealFluid}(b) for small $a_{iA}$ and $a_{iB}$ 
values, a fast heat exchange between the colloid and the fluid is expected. 
As the heat exchange is getting reduced for increasing fluid-colloid interaction strength, the 
swimming velocity is increasing. This indicates again that the temperature gradient 
between the cap and the fluid mainly determines colloid propulsion. 

The comparison of Figs. \ref{fig:MuCapRepRealFluid}(a) and \ref{fig:MuCapRepRealFluid}(b) for the stick and slip BCs have shown 
that the no-slip BCs also result in an enhancement of the swimming velocity for the low strengths of 
fluid-colloid interactions. This finding is rather counter-intuitive, since no-slip BCs lead to an 
additional friction on the colloid exerted by the fluid. Bounce-back reflections do not affect fluid
particle distribution near the colloid in comparison to specular reflections, and therefore, it is 
plausible to expect no change in heat exchange between the colloid and the fluid at least through 
the heat conduction term in Eq. (\ref{eq:cond}). The particle kinetic energy also remains conserved 
for both bounce-back and specular reflections. One should expect differences in the potential 
energy for the different collision rules. For example, bounce-back reflections may lead to a slight 
elevation of temperature ($2-5\%$) near a wall in comparison to a specular type of reflections, 
which has been found for the standard isothermal (non-energy-conserving) DPD \cite{Visser_CBC_2005}.   
A local increase of temperature near the colloid surface would reduce conductivity between the 
colloid and the fluid, which would be consistent with an increase of the swimming velocity as discussed 
above for the fluid-colloid interactions. However, currently we cannot exclude that other effects are 
present and the interplay between viscous friction and heat exchange between the colloid and the fluid 
for various parameters needs to be investigated in much more detail.  

Results for the ideal fluid have shown swimming trends qualitatively different from those for a non-ideal 
liquid. For instance, in case of bounce-back reflections the swimming velocity first decreases with 
increasing the fluid-colloid repulsive interaction strength for an ideal fluid 
(Fig. \ref{fig:MuCapRepIdealFluid}(a)), while in the corresponding case for a non-ideal fluid 
(Fig. \ref{fig:MuCapRepRealFluid}(a)) the swimming velocity is increasing when the fluid-colloid 
interactions are getting stronger. In the case of specular reflections, the trends of an increase 
of the swimming velocity with increasing the fluid-colloid interaction strength 
(Figs. \ref{fig:MuCapRepIdealFluid}(b) and \ref{fig:MuCapRepRealFluid}(b)) are similar for both 
fluid types; however, for the ideal-fluid case the swimming velocity changes its sign, which means that the
thermophoretic swimmer changes its swimming direction. From Figs. \ref{fig:MuCapRepIdealFluid}(a) 
and \ref{fig:MuCapRepIdealFluid}(b) we can also conclude that the swimmer's velocity in case of an ideal fluid becomes nearly 
independent of the repulsion strength and type of the fluid particle reflection, 
when $a_{iA} \gtrsim 100$ and $a_{iB} \gtrsim 100$, since such a repulsion strength is large enough 
to nearly push all fluid particles away from a layer of the cutoff radius ${r^{\prime}}_c$ of the fluid-colloid repulsive interaction evidenced from fluid-density 
distributions. Note that this occurs due to a high compressibility of an ideal fluid, while for 
a non-ideal fluid the layer of ${r^{\prime}}_c$ still remains populated by fluid particles. 
Therefore, the heat exchange between the colloid and the fluid is expected to be affected by the 
fluid-colloid interactions much more for ideal fluids than that for non-ideal liquids.  
Another difference between the ideal and non-ideal fluid cases is density gradients, which are 
much stronger for an ideal fluid than for a non-ideal liquid, even though temperature distributions
are not drastically different. For the case of a non-ideal fluid, the changes in density are within 
a few percent from an average fluid density, while for the ideal-fluid case the density may change 
up to 30-40\% from an average density. Thus, the fluid-density changes are more realistic in the 
non-ideal fluid case. However, currently it is not clear how these differences between the ideal 
and non-ideal fluid cases lead to different swimming behavior of the thermophoretic colloid. 
                 
A qualitative explanation for the behavior of a (homogeneous) colloidal particle in a temperature gradient has 
been sketched in Ref. \cite{Luesebrink_TPC_2012} for MPC simulations. A temperature gradient results in an 
inverse gradient of density and thus, the density around the cold side is larger than that at the hot side. 
Hence, a higher density on one side may result in a stronger interaction and lead to the colloid motion. 
A change in fluid-colloid interactions (e.g., repulsion or attraction) may invert this balance and 
force a colloid to move to an opposite direction. This idea is equivalent to having a pressure gradient 
across the colloid poles and the fluid-colloid interactions seem to provide a control for it. 
However, this argument for the generation of a pressure difference is not completely conclusive, 
because for an ideal gas in local thermodynamic equilibrium, the pressure $p=k_B T \rho$ should be 
constant due to mechanic stability. 
 
Swimming of the Janus colloid toward the hot side in Fig. \ref{fig:MuCapRepIdealFluid}(b) for a case of 
ideal fluid, specular reflections, and weak fluid-colloid interactions is consistent with this proposition. 
Note that in case of specular reflections no exchange of momentum occurs between the fluid and colloid in the 
tangential direction, and therefore, a driving force for the swimming colloid is likely to come from a  
pressure gradient across the colloid poles. Then, as we increase the repulsion between the fluid and colloid 
in Fig. \ref{fig:MuCapRepIdealFluid}(b), the pressure difference is turned around and the thermophoretic 
swimmer moves toward an opposite direction. The comparison of the specular-reflection case in 
Fig. \ref{fig:MuCapRepIdealFluid}(b) to the bounce-back BCs in Fig. \ref{fig:MuCapRepIdealFluid}(a) 
indicates that exchange of momentum between the fluid and colloid in the tangential direction also 
contributes to the swimmer propulsion, since the mobility of colloid is different for these two conditions. 
Note that no pressure differences are expected between the specular and bounce-back cases, because both 
conditions lead to the same density distributions and the same exchange of momentum between the fluid 
and colloid in the normal direction. Currently, we cannot identify the swimming effect due to the tangential 
momentum exchange, but it is clearly present in these systems. 

The application of the idea above to the case of a non-ideal fluid is not so straightforward, because the density gradients 
in this case are much smaller than in the case of an ideal fluid, as already mentioned. In fact, 
Fig. \ref{fig:MuCapRepRealFluid}(b) for specular BCs indicates that a colloid does not swim for the case of vanishing 
fluid-colloid interactions, which implies no pressure gradient across the colloid poles. As the repulsion 
between the fluid and colloid is increased in Fig. \ref{fig:MuCapRepRealFluid}(b), the swimmer starts moving toward the cold side 
indicating that a pressure difference across the colloid poles must have developed. The comparison of 
results in Fig. \ref{fig:MuCapRepRealFluid}(b) and Fig. \ref{fig:MuCapRepRealFluid}(a) for specular and 
bounce-back BCs, respectively, implies again that the exchange of momentum between the fluid and colloid 
in the tangential direction must contribute to colloid's swimming. Here, for the case of bounce-back BCs in 
Fig. \ref{fig:MuCapRepRealFluid}(a), the colloid has a non-zero swimming velocity for vanishing 
fluid-colloid interactions. Thus, the effect of tangential momentum exchange on the propulsion of a thermophoretic 
swimmer needs to be investigated further.
                            
Finally, we would like to discuss the differences between our simulations with an ideal fluid and the MPC 
simulations of a similar Janus-colloid swimmer \cite{Yang_TNS_2011,deBuyl_PSP_2013,Yang_HSS_2014}. Note that a direct 
comparison has not been intended. A seeming dissimilarity is the dependence 
of swimming velocity on the fluid-colloid interaction strength. In the present simulations the thermophoretic 
colloid swims toward the cold side if we increase the repulsive strength of fluid-colloid interactions, while 
in Refs. \cite{Yang_TNS_2011,Yang_HSS_2014} repulsive interactions between a colloid and a fluid result in 
the motion toward the hot side. A closer look at the details of the simulation setups reveals that the 
fluid-colloid interactions in these two studies may have a different meaning. In our setup, the repulsive 
interactions directly affect heat exchange between the colloid and the fluid, since larger distances 
between them reduce the exchange of heat. In Refs. \cite{Yang_TNS_2011,Yang_HSS_2014} the temperature 
in a thin layer near the colloid is controlled, and thus, the repulsive interactions affect the number of 
particles to be thermalized in this layer. Hence, in the present simulations repulsive interactions 
affect conductivity between the colloid and the fluid, while in Refs. \cite{Yang_TNS_2011,Yang_HSS_2014}
such conductivity effects are omitted, which is equivalent to a very high colloid-fluid conductivity such that 
a thin layer of fluid particles near the colloid receives heat instantly. Another difference between 
the two simulation setups is temperature control. In our simulations, generated heat is taken away at 
the cold side of the colloid or far away from the colloid, while in Refs. \cite{Yang_TNS_2011,Yang_HSS_2014}
the excess heat is taken away uniformly from the whole fluid. This may result in different temperature 
distributions around the colloid affecting its swimming behavior. Finally, there exist a fundamental 
difference between the simulation methods. In MPC, heat exchange and temperature gradients are sustained 
only through the kinetic energy of fluid particles, while in eDPD an internal energy is simulated 
explicitly. In fact, the internal energy is much larger than the contributions from kinetic and 
potential energies, since $C_v \gg 1$. Simulations with $C_v \approx 1$ appear not to be stable in eDPD, since 
then there is a chance that internal temperature of a particle may become negative, for instance, due 
to the random conductivity term in Eq. (\ref{eq:cond}). The discussed reasons do not allow us to make 
a detailed comparison, which would require more consistent setups.                        

In conclusion, we have presented simulations of the dynamics of a thermophoretic colloid for different 
fluid-colloid interactions and temperature controls. Different temperature-control strategies have a 
minor effect on the colloid swimming velocity. The fluid-colloid interactions have a strong effect on 
the colloid behavior and directly affect heat exchange between the colloid surface and the fluid. Our 
results show that a reduction of the heat exchange leads to an enhancement of colloid's thermophoretic 
mobility, since larger temperature gradients near the colloid surface are formed. The flow-field generated 
by the colloid appears to be dominated by a source-dipole contribution in agreement with the recent 
theoretical \cite{Bickel_FPV_2013} and simulation \cite{Yang_TNS_2011,Yang_HSS_2014} predictions. 
However, the differences in colloid's mobility between the cases with non-ideal and ideal fluids and in 
comparison to the MPC simulations \cite{Yang_TNS_2011,Yang_HSS_2014} are yet to be understood. We 
hope that this work will generate further efforts and discussions in this area of research.

\section*{Acknowledgments}
We would like to thank Marisol Ripoll for discussions.
Dmitry A. Fedosov acknowledges funding by the Alexander von Humboldt Foundation.
We also gratefully acknowledge a CPU time grant by the J\"ulich Supercomputing Center.
Partial financial support by the Deutsche Forschungsgemeinschaft via the priority program "Microswimmers" 
(SPP 1726) is thankfully acknowledged.


\balance


\providecommand*{\mcitethebibliography}{\thebibliography}
\csname @ifundefined\endcsname{endmcitethebibliography}
{\let\endmcitethebibliography\endthebibliography}{}

\end{document}